\documentclass[journal]{vgtc}         

\ifpdf
  \pdfoutput=1\relax                   
  \usepackage{graphicx}                
  \DeclareGraphicsExtensions{.pdf,.png,.jpg,.jpeg} 
\else
  \usepackage{graphicx}                
  \DeclareGraphicsExtensions{.eps}     
\fi%

\graphicspath{{figures/}{pictures/}{images/}{./}} 

\usepackage{microtype}                 
\PassOptionsToPackage{warn}{textcomp}  
\usepackage{textcomp}                  
\usepackage{mathptmx}                  
\usepackage{times}                     
\usepackage{cite}                      
\usepackage{tabu}                      
\usepackage{booktabs}     
\usepackage{booktabs}                  
\usepackage{appendix}                  

\usepackage{amsmath}

\usepackage{color} 
\usepackage{colortbl}
\usepackage{cases}
\definecolor{light-gray}{gray}{0.6}
\definecolor{lavender}{rgb}{0.5,0.5,1.0}

\def\_{\rule{.3em}{.15ex}}      

\newcommand{\hot}[1]{{\color{black} #1}}
\newcommand{\comp}[1]{{\color{black}#1}}

\newenvironment{myitemize}{
\begin{itemize}
  \setlength{\itemsep}{1pt}
  \setlength{\parskip}{0pt}
  \setlength{\parsep}{0pt}}{\end{itemize}
}

\usepackage{lipsum}
\usepackage{caption}
\usepackage{tikz}
\definecolor{SomeColor}{RGB}{0,150,255}

\usepackage{subcaption}
\usepackage{xspace}
\usepackage{amssymb}
\usepackage{xcolor}
\usepackage{hyperref}
\usepackage[capitalise]{cleveref}
\usetikzlibrary{calc}
\newcommand{\ours}{flow encoder\xspace}
\definecolor{background}{HTML}{1f77b4}
\definecolor{piggypink}{HTML}{fddde6}
\definecolor{orangepeel}{HTML}{ff9f00}
\definecolor{purpleheart}{HTML}{69359c}
\definecolor{radicalred}{HTML}{ff355e}
\definecolor{green}{HTML}{00ff00}
\definecolor{purple}{HTML}{800080}

\onlineid{1349}

\vgtccategory{Research}
\vgtcpapertype{Technique}

\title{Automatic Semantic Alignment of Flow Pattern Representations for Exploration with Large Language Models}

\author{Weihan Zhang, and Jun Tao, \textit{Member, IEEE}}
\authorfooter{
\item
 Weihan Zhang is with the School of Computer Science and Engineering, Sun Yat-sen University. E-mail: zhangwh79@mail2.sysu.edu.cn.
\item
 Jun Tao is with the School of Computer Science and Engineering, Sun Yat-sen University, Guangdong Province Key Laboratory of Computational Science, and National Supercomputer Center in Guangzhou. E-mail: taoj23@mail.sysu.edu.cn. He is the corresponding author.
}

\abstract{
Explorative flow visualization allows domain experts to analyze complex flow structures by interactively investigating flow patterns. However, traditional visual interfaces often rely on specialized graphical representations and interactions, which require additional effort to learn and use. Natural language interaction offers a more intuitive alternative, but teaching machines to recognize diverse scientific concepts and extract corresponding structures from flow data poses a significant challenge. In this paper, we introduce an automated framework that aligns flow pattern representations with the semantic space of large language models (LLMs), eliminating the need for manual labeling. Our approach encodes streamline segments using a denoising autoencoder and maps the generated flow pattern representations to LLM embeddings via a projector layer. This alignment empowers semantic matching between textual embeddings and flow representations through an attention mechanism, enabling the extraction of corresponding flow patterns based on textual descriptions. To enhance accessibility, we develop an interactive interface that allows users to query and visualize flow structures using natural language. Through case studies, we demonstrate the effectiveness of our framework in enabling intuitive and intelligent flow exploration.

} 


\begin{document}



\maketitle
\section{Introduction}

Flow visualization has been crucial in analyzing and exploring flow patterns in the scientific visualization community for decades. It effectively conveys information about relevant flow structure visualization results in a manner tailored to specific domain needs.
\comp{Early flow visualization methods relied on predefined criteria to generate results. For instance, streamline placement strategies aimed at even spacing~\cite{jobard1997creating} or maximizing information content~\cite{xu2010information} were common. Other techniques focused on visualizing specific structures~\cite{theisel2003saddle,sadlo2004vorticity,zafar2023hairpin}. These approaches typically depended on expert interpretation or collaboration with domain specialists to determine target features. However, fixed-criteria methods often struggle to address domain-specific needs or generalize beyond predefined feature types, requiring significant effort to extend.

To increase flexibility, interactive visualization techniques have emerged, allowing users to explore and customize results based on their analysis goals. These methods support user-defined streamlines via graph-based interfaces~\cite{han2018flownet,tao2017semantic}, pattern queries~\cite{wang2017stream}, predicates~\cite{salzbrunn2006predicates}, and tangible interfaces~\cite{jackson2013lightweight}, with the latter being especially prominent.
Despite their flexibility, interactive techniques often involve a steep learning curve and considerable cognitive effort. While tangible interfaces offer intuitive feedback, their limited degrees of freedom constrain the formulation of complex queries. FlowNL~\cite{huang2022flownl} introduced a natural language interface for querying visualizations, translating user input into a declarative language for flow data filtering. However, it still depends on manual labeling of flow patterns.
Recent advances in large language models (LLMs) have shown promise in bridging natural language and complex data representations. Nevertheless, enabling LLMs to understand flow structures remains a significant challenge.}

To tackle these challenges, we propose a framework aimed at aligning flow representations with natural language semantic features automatically. Our framework eliminates the need for manually specifying the correspondence between flow patterns and their textual descriptions.
The framework consists of three main components: an \emph{autoencoder} representing the flow pattern, a \emph{semantic alignment module} aligning pattern representations with text embedding, and a \emph{query module} matching pattern representations with text queries. 
\hot{Before explaining the components, we should note that while our framework takes spatial curves as input, which is applicable to both streamlines in steady fields and pathlines in unsteady fields, this work focuses on streamlines and steady fields. Its effectiveness for unsteady fields remains to be evaluated.}

\textbf{Flow pattern representation.} 
These flow structures can typically be reflected locally by sample points along streamlines. Therefore, to cover and represent a broad range of flow patterns, we sample streamline segments directly from the traced streamlines for pattern representation. The representation starts a handcrafted transformation that converts a sampled streamline segment to a matrix containing the pairwise distance of the sample points. This distance matrix provides a concise representation of spatial relationships among points along streamline segments, thus capturing the corresponding flow pattern. The distance matrix effectively encodes the shape of a streamline segment, which is invariant under rigid transformations.
\hot{In order to capture diverse flow patterns with a unified latent representation space, we feed the distance matrices to a denoising autoencoder, which produces discriminative vectors in the embedding space.} 
The resulting latent vectors obtained from the autoencoder are then used as flow representations, providing a compact, effective encoding of the flow characteristics embedded within each streamline segment.

\textbf{Connecting pattern representations to semantics.}
Our approach is inspired by recent advancements in multimodal large language models in various domains~~\cite{dai2023instructblip,liu2024llava,xu2024pointllm}, which commonly encompass alignment and instruction tuning that teach LLMs to recognise objects in different forms. 
In our case, we design a projector that maps flow pattern representations into a semantic space aligned consistently with the textual embeddings of LLMs.
To train the projector for alignment, we further propose an automatic text-flow instruction-following data generation scheme. This data is used for teaching the model to interpret flow data and follow user instructions. To avoid costly and labour-intensive manual data compilation, we use GPT-4o for automated data collection, incorporating streamline images and textual descriptions. 
Intuitively, the training data connects streamline images and meaningful semantics, and the training process further connects flow representations and semantics. 
These steps jointly enable LLMs to understand and process flow data effectively.

\textbf{Representation matching from text.} The semantic matching between textual descriptions and flow data enables the automatic generation of visualizations, removing the need to rely on predefined or fixed criteria. Since we have already constructed representative encoded features for streamline segments, matching flow features with textual embeddings becomes straightforward. While similarity-based matching is common practice, it requires alignment between the embeddings of text and flow data in a shared semantic embedding space. To accomplish this, we introduce a lightweight semantic matching model that leverages a cross-modal attention mechanism to learn correspondences across text and flow modalities effectively. 
This allows us to use natural language to match similar flow representations. To improve accessibility, we have also developed an interactive interface that enables users to query and visualize flow structures intuitively through natural language. 

In summary, the contributions of this paper can be described as follows:
\begin{myitemize}
\item We propose a simple yet effective flow pattern representation framework based on a denoising autoencoder, which converts streamline segments into distance matrices for unsupervised reconstruction. This generates pattern representations that are invariant under rigid transformations.

\item We propose an automatic semantic alignment framework that assigns meaningful semantics to the flow representations, by leveraging the knowledge embedded in LLMs. The framework allows flow patterns to be described and queried in the textual format without manual labelling.

\item We develop an interactive interface for visualizing specific flow patterns identified from conversations, demonstrating our framework's effectiveness and accessibility.

\end{myitemize}

\section{Related Work}

In this section, we discuss various approaches proposed to extract, match, and understand flow patterns based on streamlines. We then introduce existing work on visualization driven by natural language queries. Finally, we review recent work on multimodal large language models, whose frameworks closely resemble our proposed approach.

\textbf{Streamline-based flow pattern.}
Geometric flow visualization has been extensively studied over the past few 
decades~\cite{mcloughlin2010flowsurvey}, with numerous methods proposed to extract, compare, and interpret streamline-based flow patterns. A significant body of research has focused on developing similarity metrics for streamlines, aiming to quantify the resemblance between different flow structures and enhance the accuracy of flow pattern analysis. 
Chen~et~al.~\cite{chen2007similarity} defined the streamline distance based on point-wise distances within sliding windows.
Wei~et~al.~\cite{wei2010sketch} represented streamlines as sequences of curvature and torsion value pairs. The distance between two streamlines was determined by the edit distance between the corresponding sequences.
Rössl~et~al.~\cite{rossl2012embedding} embedded streamlines as 3D points based on the Hausdorff distance for exploration.
Tao~et~al.~\cite{tao2014flowstring} advocated the use of Procrustes distance to achieve invariance to rigid transformations. Streamlines were sampled based on cumulative curvature to ensure scale invariance. This method was later extended to support multi-level features and multiple datasets~\cite{tao2015vocabulary}.
Oeltze~et~al.~\cite{oeltze2014blood} investigated the impact of distance metrics and clustering techniques on blood flow clustering, considering both point-wise distances and attribute distances.
Wang~et~al.~\cite{wang2017stream} enabled pattern queries across multiple streamlines by first identifying candidate streamline segments and then aligning the query pattern with the candidate segments.

Beyond these approaches, other methods involve generating streamline descriptors to measure similarities in an abstract feature space. This approach enables a more flexible and generalized representation of flow structures.
Descriptors are used to describe the shapes of 3D curves or geometric objects. Early methods appeared in computer graphics, collecting sampled information to form histograms as descriptors. 
For example, Osada~et~al.~\cite{osada2002shape} proposed a random descriptor called the D2 descriptor, which is essentially a histogram of distances between sampled point pairs.
For flow pattern analysis, researchers often consider attributes beyond sampled point distances~\cite{shi2009path}. 
Brun~et~al.~\cite{brun2004clustering} mapped fiber traces to a high dimensional feature space, where fibers were clustered through normalized cuts. 
McLoughlin~et~al.~\cite{mcloughlin2013similarity} described streamlines using curvature histograms and computed their similarity with $\chi^2$ tests. 
Similarly, Lu~et~al.~\cite{lu2013exploring} designed their descriptors as 2D histograms of curvature, with each row corresponding to a curvature range and each column associated with a streamline segment. 
Li~et~al.~\cite{li2014streamline} proposed a bag-of-features-based signature that preserved the order of points better than histograms, considering the streamline attributes and spatial information. 
FlowNet~\cite{han2018flownet} represented streamlines and flow surfaces as binary volumes, enabling the direct application of standard 3D convolutional network architectures to learn spatial patterns and distributions of streamlines. 

\textbf{Natural language interaction.}
Natural language interaction has been explored in information visualization for nearly a decade. Most relevant methods follow a common framework: they parse natural language queries and translate them into an intermediate representation, such as explicit commands~\cite{sun2010articulate}, SQL queries~\cite{dhamdhere2017analyza}, VisFlow functions~\cite{yu2019flowsense}, or declarative specifications~\cite{narechania2020nl4dv}. The visualization is then generated based on these intermediate commands, enabling users to interact with data more intuitively. Luo~et~al.~\cite{luo2021nl2vis} followed a similar framework but introduced a transformer-based model to enhance the translation process. Additionally, several interfaces incorporate interactive dialogues between users and the system to facilitate more dynamic and responsive interactions. Notable examples include Articulate 2~\cite{kumar2016articulate2}, Eviza~\cite{setlur2016eviza}, and Evizeon~\cite{hoque2017applying}, which enable users to refine queries, receive clarifications, and iteratively explore data through natural language conversations. 
All of the works above support natural language query processing for visualization. The most closely related work to ours is FlowNL~\cite{huang2022flownl}, which translates natural language descriptions into the declarative language to generate flow visualizations.

\textbf{Multi-modal large language models.}
Multimodal Large Language Models aim to understand and interpret a wide range of information beyond pure text-based data, including but not limited to images, audio, and more \cite{yin2023survey}. To enable LLMs to understand specific modal data, end-to-end training strategies are commonly employed \cite{li2022blip}. However, this approach poses challenges, including the need for large-scale data and the substantial cost of training from scratch. The better paradigm builds on pre-trained LLMs and modal encoders \cite{alayrac2022flamingo,dai2023instructblip,li2023blip2,liu2024llava}. This strategy typically involves a two-stage process: aligning the modal encoder to the feature space of the LLM, followed by instruction-based alignment. 
Flamingo \cite{alayrac2022flamingo} uses few-shot learning to handle interleaved image and text data, achieving excellent performance on specific tasks.
InstructBLIP \cite{dai2023instructblip} leverages pre-trained vision encoders and language models, using cross-attention to enhance the model’s multimodal understanding capabilities.
Li~et~al.~\cite{li2023blip2} proposed a general and efficient strategy for vision-language models by guiding vision-language pretraining from off-the-shelf frozen pre-trained image encoders and frozen large language models.
Liu~et~al.~\cite{liu2024llava} proposed an end-to-end trained large multimodal model that connects a vision encoder and an LLM for general-purpose visual and language understanding, achieved through instruction tuning on generated data. 
Similar approaches have also been explored in visualization. For instance, Zeng~et~al. ~\cite{zeng2024mllm4chartQA} employed a pre-trained visual encoder and utilized a projection layer to map charts into the semantic space of LLMs. This integration significantly improved performance on chart-based question-answering tasks, demonstrating the potential of aligning visual representations with language models for enhanced reasoning and understanding.

\textbf{Comparison with the existing works.}
Several semantic-based interaction approaches have been developed for flow structure representation. Among these techniques, semantic flow graphs~\cite{tao2017semantic}, streamline predicates~\cite{salzbrunn2006predicates}, and IGScript~\cite{liu2021IGScript} identify streamlines based on conventional attributes. FlowNL~\cite{huang2022flownl} supports segment-level queries using a special attribute (i.e., latent representations), but it relies on manual effort to label the latent space for language-based queries. Our approach is similar to existing approaches as it queries streamline segments based on their representation. 
\hot{However, it improves the existing approaches in automatically aligning the flow representation with text embedding and LLMs' knowledge. This allows the flow patterns to be queried by text in various domains with minimal manual intervention. 
}
\section{Our Framework}

\begin{figure}
    \centering
    \includegraphics[width=1\linewidth]{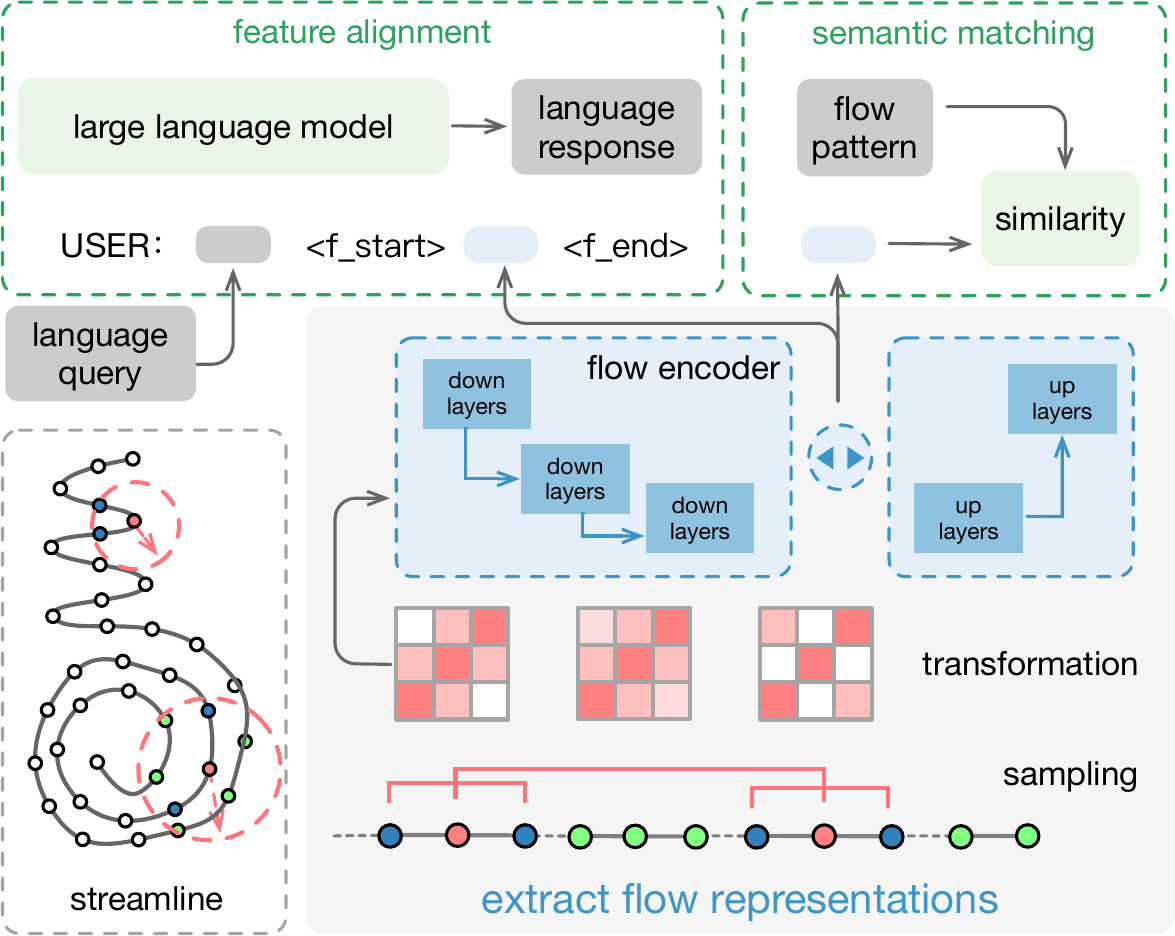}
    \caption{The illustration depicts our automatic semantic alignment framework. We begin by sampling streamline segments and transforming them into a suitable format for our \ours. The latent vectors extracted from the flow encoder serve as the flow representations. In the feature alignment phase, we integrate textual input with these flow representations and feed them into LLMs to generate context-aware textual responses. Finally, in the semantic matching phase, we encode flow patterns into text embeddings, enabling the retrieval of similar flow representations based on semantic similarity.}
    \label{fig:framework}
\end{figure}

Our automatic semantic alignment framework consists of three main components: \emph{flow pattern representation}, \emph{semantic alignment}, and \emph{semantic matching}, as illustrated in~\cref{fig:framework}. The flow pattern representation is achieved by sampling streamline segments and encoding them as latent vectors, which ensures rigid invariant flow representations. Then, the semantic alignment aligns these flow representations with natural language by mapping them into a shared semantic space, enabling LLMs to comprehend and interact with flow data. Finally, the semantic matching module projects flow representations into the semantic space and computes their similarity to textual queries.

\subsection{Flow Representation}
Streamlines are commonly used to visualize flow patterns. However, due to the influence of the streamline tracing process, they often encompass diverse flow patterns. Therefore, we first sample segments from the streamline to capture the flow patterns they encompass. We aim to extract these flow patterns from streamline segments while addressing common challenges related to inexactness and limited robustness to shape variations. 
Additionally, we use a \ours to encode these extracted streamline segments into vectors, enabling their alignment with the semantic space.
In this section, we introduce the process of sampling segments from streamlines and encoding them into vectors.

\textbf{Sampling from streamlines.}
\hot{Flow structures exhibit varying spatial scales, which necessitates the use of streamline segments of different lengths during sampling. To capture large-scale patterns, we sample longer segments along streamlines. Subsequently, we decrease the segment length to cover finer-grained structures. This hierarchical sampling strategy ensures comprehensive coverage across scales: it preserves global features in long streamlines by avoiding overly short segments, while enabling finer analysis in shorter streamlines by extracting multiple overlapping segments.}
Following this strategy, we obtain streamline segments of varying lengths. We then apply transformation and normalization to these segments to facilitate a unified neural network for extracting flow representations.

\textbf{Transformation and normalization.}
To effectively capture the patterns of streamline segments, we employ a novel descriptor inspired by visibility graphs~\cite{lacasa2008visibilitygraph} for time series data exploration. Specifically, we transform streamline segments into distance matrices derived from their control points. These matrices remain rigid-invariant, ensuring that they faithfully preserve flow patterns while serving as more suitable inputs for our \ours.
For efficient computation and normalization, we generate these distance matrices by uniformly sampling a fixed number of control points from each streamline segment. However, the current distance matrix is influenced by the arc length of the streamline segment, which significantly affects the distances between sampled points. To mitigate this effect and ensure consistency, we normalize the distance matrices by dividing all values by the arc length of the corresponding streamline segment.

\textbf{flow representations from \ours.}
Since our distance matrices inherently capture the shape of the streamline segments while maintaining rigid invariance, they provide a strong foundation for learning flow representations using neural networks. Given the absence of labeled data, we adopt a self-supervised learning approach to extract meaningful flow features.
Denoising Autoencoders (DAEs) have proven to be effective self-supervised learners~\cite{vincent2008dae,xiang2023ddae}. We leverage the DAE as our \ours to extract high-quality flow representations from streamline segments. Inspired by prior work~\cite{xiang2023ddae}, we incorporate diffusion models into the denoising process by injecting noise into the distance matrices. This encourages our \ours to learn robust and discriminative flow representations.
Although existing methods~\cite{preechakul2022diffusion,zhang2022unsupervised} employ auxiliary encoders for representation learning, they often neglect the discriminability of features. Meanwhile, approaches based on modified diffusion frameworks~\cite{abstreiter2021diffusion,mittal2022points} aim to learn linearly separable features but fail to achieve optimal performance. Recent studies~\cite{xiang2023ddae,chen2024ldm} have demonstrated that while diffusion models are primarily designed for generative tasks, their reverse denoising process can be highly effective for representation learning.

Building on this insight, we adopt the diffusion model framework to train our \ours, ensuring the generation of high-quality and robust flow representations.
Diffusion models~\cite{ho2020ddpm,karras2022edm,song2020sde} apply Gaussian noise at different levels to corrupt the data $x_0$. The noise at different levels is controlled by the specified time steps $t=1,..., T$. The corrupted data, i.e., the noisy data, is defined as $q(x_t \mid x_0) = \mathcal{N}(\alpha_t x_0, \sigma_t^2 \mathbf{I})
$, where $\alpha_t$ and $\sigma_t$ are the hyperparameters controlling the scale of the original data and noise fusion. Given a specific $t$, a noised version of $x_0$ at the $t$ time step can be computed by sampling from $\epsilon \sim \mathcal{N}(0, \mathbf{I})$ and taking:

\begin{equation}
    x_t = \alpha_t x_0 + \sigma_t \epsilon.
\end{equation}

\noindent The training of diffusion models typically aims to predict the noise from the noised data in order to reverse the diffusion process. Specifically, a random $x_t$ is first sampled from $\mathcal{N}(0, \mathbf{I})$, and the model is trained to infer: $x_{t-1} \sim q(x_{t-1} \mid x_t)$. Diffusion models use neural networks to learn the transitions with $p_{\theta}(x_{t-1} \mid x_t) = \mathcal{N}(\mu_\theta(x_t, t), \Sigma_t^2 \mathbf{I})$, whose mean $\mu_\theta(x_t, t)$ is obtained through the network, and the variance $\Sigma_t^2$ is a constant value.

\textbf{Implementation of \ours.}
We should first recognise a critical issue: we need to adjust the training objective of the network to predict the original input data directly. In our task, we aim for the network to directly reconstruct the distance matrices of streamline segments, thus forming an autoencoder-based structure. By simplifying the maximum likelihood objective, the network should learn the denoising autoencoder objective~\cite{karras2022edm}:

\begin{equation}
    \text{Loss} = \|D_\theta(x_t, t) - x_0\|^2,
\end{equation}

\noindent where $D_\theta(x_t, t)$ is a function that transforms $x_t$ and $\mu_\theta(x_t, t)$ back to the original input. Therefore, we use a neural network to approximate this denoising process, and this network can be viewed as the DAEs that accept different noise-scaled versions of the data. We can directly use the noise schedule from DDPM~\cite{ho2020ddpm}, following the work~\cite{xiang2023ddae}. The parameterization strategy proposed by DDPM is actually variance preserving, where $\alpha_t = \sqrt{\prod_{i=1}^{t} (1 - \beta_i)}$ and $\alpha_t^2+\sigma_t^2=1$. $\beta_i$ is obtained from a linear schedule. The original DDPM is designed for generative tasks and is thus used to predict noise. However, according to work~\cite{chen2024ldm}, the capability of DAEs can also be derived from predicting noise: $D_\theta(x_t, t)=\frac{x_t - \sigma_t \epsilon_\theta(x_t, t)}{\alpha_t}$, where $\epsilon_\theta$ is the network for predicting noise. 

We choose UNet as the approximation network for denoising in our \ours. The distance matrices are used as $x_0$, which is first noised to $x_t$. The network then predicts $x_0$ to generate flow pattern representations. We use the encoder structure of the trained network to encode the distance matrices of streamline segments to latent vectors, which are used for alignment to language models.

\subsection{Semantic Alignment to Large Language Models }
\hot{The semantic alignment module aligns flow representations of streamline segments with the input space of LLMs. This addresses the semantic mismatch between the two modalities: LLMs are trained on sequences of token embeddings derived from discrete language tokens, whereas flow representations are continuous vectors produced by \ours. Due to the differing structure and semantics of these representations, we introduce a projector network to map flow features into the LLM's embedding space, enabling effective cross-modal integration.}
In addition, our semantic alignment targets an automatic generation of training data that allows LLMs to differentiate different flow patterns and associate textual descriptions with underlying flow structures, requiring training datasets for instruction fine-tuning LLMs.

\begin{figure}[h]
    \centering
    \includegraphics[width=1\linewidth]{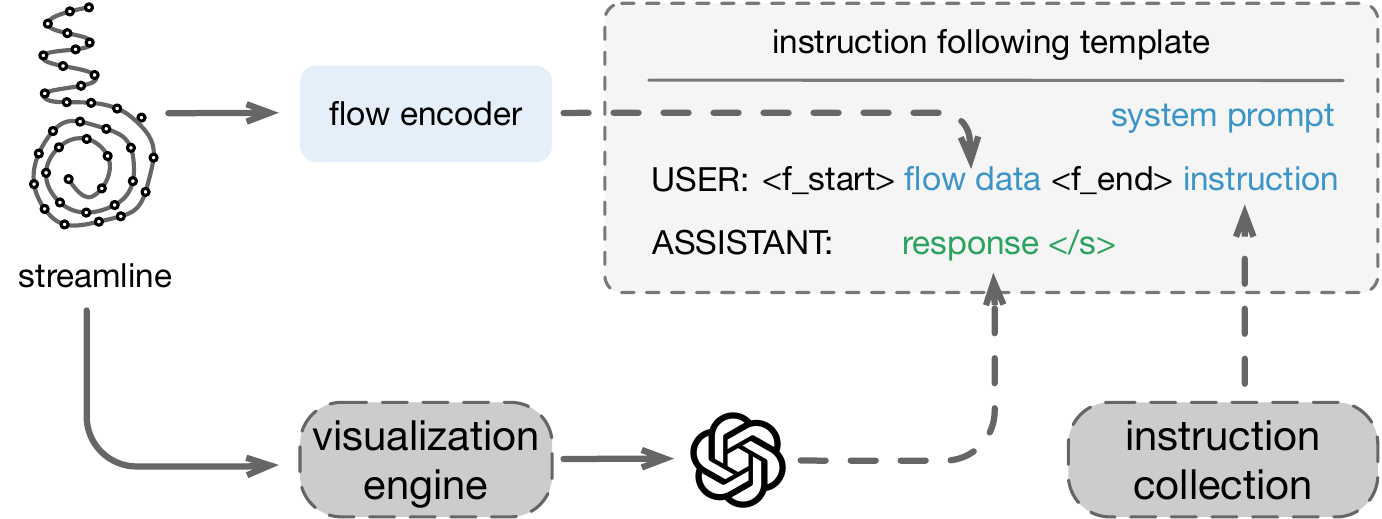}
    \caption{The generation pipeline of text-flow instruction-following data. The rendered streamline results are prompted into GPT, combined with instruction queries, to generate corresponding textual responses. Additionally, flow representations derived from \ours serve as multimodal flow data.}
    \label{fig:data}
\end{figure}

\textbf{Text-flow instruction-following data.}
A longstanding challenge in developing end-to-end multimodal models is acquiring high-quality multimodal instruction-following data. Such data is essential for effective representation learning, playing a crucial role in aligning the latent space and ensuring coherent multimodal understanding and reasoning~\cite{alayrac2022flamingo,dai2023instructblip,liu2024llava,xu2024pointllm}. However, manual labelling of scientific data is both costly and labour-intensive.
To address this challenge, we adopt the approach proposed in~\cite{liu2024llava,xu2024pointllm}, leveraging GPT-4o to automatically generate instruction-following data. As illustrated in~\cref{fig:data}, the generated data follows a uniform instruction following template, consisting of descriptive instructions for streamlines and complex reasoning instructions. The descriptive instructions are designed using diverse question structures, focusing on characterizing the shape features of streamlines. In contrast, the reasoning instructions explore flow dynamics and potential flow patterns. 
\hot{To ensure instruction diversity, we define a collection of prompt templates and randomly sample these prompts from a predefined instruction collection (please refer to the supplementary material for details).
During data collection, we prompt GPT-4o with multi-view information, enabling it to generate comprehensive textual responses. We manually perform a brief check to ensure the generated responses align with the intended semantics. Specifically, we randomly sample and inspect approximately $20\%$ of the generated responses for each dataset, which typically takes around 10–15 minutes.}
For each flow field, we collect approximately five hundred text-flow instruction-following samples. This approach enables the automatic collection of multimodal data, significantly reducing the need for manual annotation. During LLM fine-tuning, these data samples help the model learn the correspondence between input flow representations and textual descriptions in an automated manner, thereby enhancing its ability to perceive and interpret flow structures.

\begin{figure}
    \centering
    \includegraphics[width=1\linewidth]{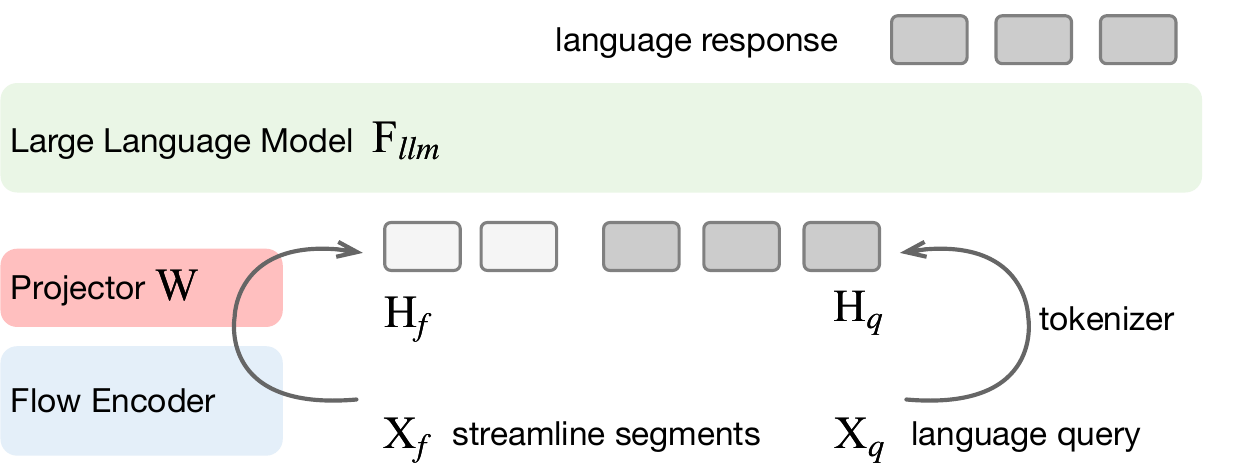}
    \caption{The illustration of our strategy for aligning streamline segment features to the LLM's latent space. The \ours extracts the flow representations from streamline segments, and the projector maps them into the latent space of the LLM backbone. The LLM then generates the predicted tokens as the output.}
    \label{fig:LLM}
\end{figure}

\textbf{Model structure.}
As shown in~\cref{fig:LLM}, the framework is a generative model that aims to complete multi-modal sentences containing both streamline segments and texts. The model consists of three main components: a pre-trained \ours, a projector, and a pre-trained LLM backbone. Our framework is designed to support the use of any encoder for obtaining flow representations. As an example, we employ a flow encoder; however, it is not inherently tied to our semantic alignment framework.

The \ours takes the distance matrices of streamline segments $\textbf{X}_f$ as input and encodes them into features. The projector $W$ is essentially an MLP network that further maps the features extracted by the \ours into $\textbf{H}_f$. This process can be denoted as:
\begin{equation}
    \textbf{H}_f=\textbf{W} \cdot \text{\ours}(\textbf{X}_f).
    \label{eq:projection}
\end{equation}

\noindent The text query instructions $\textbf{X}_q$ are mapped into the LLM's latent space via the tokenizer, which is equivalent to the process in~\cref{eq:projection}. Since the tokenizer is essentially aligned with the LLM, $\textbf{H}_q$ can be treated as input tokens for the LLM.
It is important to note that the alignment process does not simply map flow representations to specific textual tokens. We employ a projector to mimic the embedding space of the LLM. In practice, the projected vectors are concatenated with the LLM’s input embeddings, ensuring an integration of flow representations into the model’s latent space. This approach allows for a more structured fusion of multi-modal information, facilitating improved downstream understanding and reasoning. To distinguish between flow representations and textual modalities, we introduce additional special tokens, $<\text{f\_start}>$ and $<\text{f\_end}>$. The projected vectors are placed between these two special tokens, effectively segmenting the flow representation from textual input within the LLM’s embedding space. 

The LLM backbone $\textbf{F}_{llm}$ is a specific decoder-only large language model~\cite{dubey2024llama3}, which takes a series of tokens as input and outputs a series of tokens. Our input tokens contain a mixture of streamline segment features and text instructions, which is denoted as $Z = (z_1, z_2, \dots, z_k) $. Using the attention mechanism, the $\textbf{F}_{llm}$ is able to understand the contextual relationships between different types of tokens, enabling it to generate responses.
In this process, we leverage the attention mechanism to enable the LLM to automatically extract semantically relevant features from the flow representations~\cite{xu2024pointllm}. By dynamically attending to different parts of the projected vectors, the model can learn to identify meaningful correlations between the flow data and textual descriptions. Although our training data does not provide a comprehensive description of all possible flow patterns, the fine-tuning process enables the LLM to align flow patterns with semantically similar textual representations. Leveraging its pre-trained knowledge, the LLM can develop an implicit understanding of these concepts, allowing it to generalize beyond the provided data. Through multiple decoding, the LLM generates response text by predicting the next token, as follows:

\begin{equation}
    \hat{z}_i=\textbf{F}_{llm}(Z_{<i}).
    \label{eq:next_token}
\end{equation}

\noindent The output of the LLM backbone $\textbf{F}_{llm}$ is a sequence of predicted tokens. The $i$-th predicted token $\hat{z}_i$ is calculated based on all previous tokens:

\begin{equation}
    \tilde{z}_i = \arg\max_{w \in \text{vocab}} f_{\text{vocab}}(\hat{z}_i)[w].
    \label{eq:generate}
\end{equation}

\noindent To generate the language response, each $\hat{z}_i$ is passed through a final linear layer followed by a softmax operation, mapping the hidden states into a probability distribution over the vocabulary. As demonstrated in~\cref{eq:generate}, this final layer is defined as $f_{\text{vocab}}$. Its dimension depends on the vocabulary size, where additional special tokens need to be included. The final prediction $\hat{z}_i$ represents the word corresponding to the $i$-th token, which is selected based on its highest probability from the vocabulary. 

\textbf{Instruction tuning.}
We train our model by minimizing the negative log-likelihood of the next token at each position. Our loss function is computed only on the model's response tokens, and we employ an end-to-end training approach. This allows our model to effectively integrate both the streamline and textual modalities. Our training process is divided into two stages, each corresponding to different data types and focusing on distinct training objectives. 

During the first stage of training, also referred to as the feature alignment phase, our objective is to shift the mapping space of the projector closer to the semantic space of the LLM backbone. We freeze the parameters of the \ours and LLM, and only train the parameters of the MLP projector. We use description instruction data for training, aiming to align the encoding space of the streamline with the textual token space effectively. 
In the second stage, we continue to freeze the parameters of the \ours, while jointly training the projector and the LLM backbone. To preserve the general capabilities of the LLM while adapting it to understand streamline data, we employ LoRA fine-tuning. LoRA fine-tuning efficiently updates parameters by introducing low-rank matrices, allowing the model to capture task-specific knowledge even with limited data. During this process, we use complex reasoning instruction data for training, helping the model develop the ability to understand and respond to complex instructions. This process further trains the projector, enabling the LLM to better understand streamline data.

\begin{figure}
    \centering
    \includegraphics[width=1\linewidth]{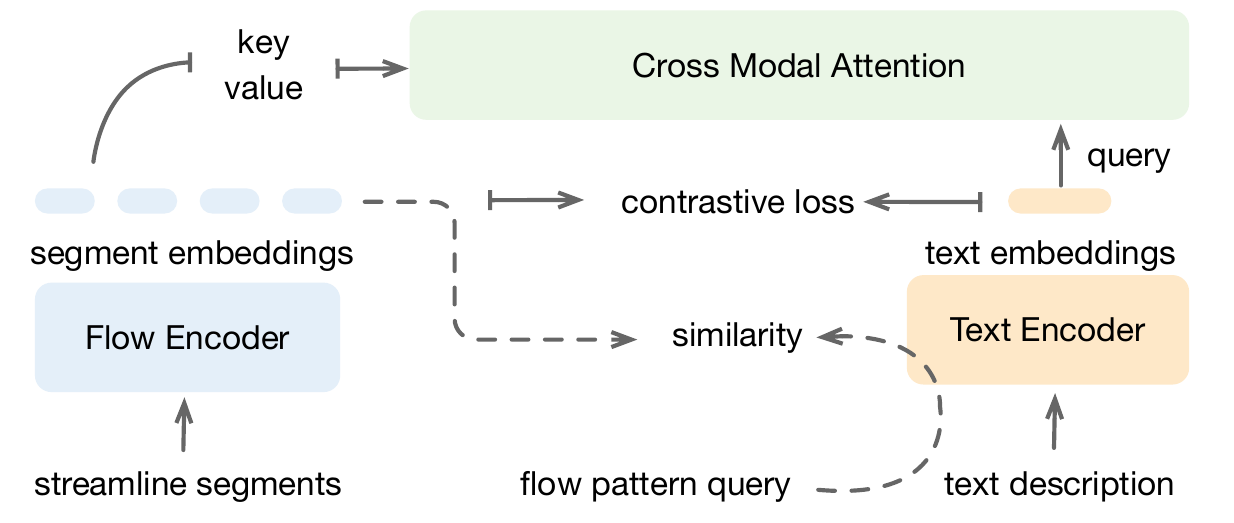}
    \caption{The illustration of our semantic matching method. The text encoder utilizes the pre-trained embedding model, while the streamline segment encoder employs our~\ours. During modality fusion, the text embedding serves as the query, and the streamline segment embeddings act as the key and value. Cross-modal attention is leveraged to learn their associations.}
    \label{fig:matcher}
\end{figure}

\subsection{Semantic Matching}
The semantic matching identifies semantically similar streamline segments based on textual queries. These queries typically describe specific flow patterns in a textual manner, such as ``spiral-like flow" or ``turbulence". We first discuss the potential algorithmic choices and then introduce our solution.

\textbf{Alternative algorithmic choices.} A straightforward solution approach is to compute the similarity between textual queries and projected flow vectors, given that both are mapped into a shared space during the feature alignment phase. However, this method has notable limitations due to the incomplete overlap between the positional distributions of textual queries and projected vectors within the common space. As a result, each query typically corresponds to only a limited subset of the nearest projected vectors, potentially overlooking relevant flow patterns. Furthermore, the query embedding's reliance on the LLM tokenizer introduces an additional constraint, making it challenging to generate a unified embedding that fully encapsulates the textual query.
An alternative approach involves directly generating matching flow vectors from textual queries and then retrieving the nearest streamline segments. This method can be implemented using autoregressive generation techniques\cite{ramesh2021dalle}. However, it introduces significant challenges, particularly regarding vector quantization requirements. Given the limited availability of flow data, effectively training such quantization is difficult. Moreover, since flow representations exist in a continuous latent space, discretizing them through vector quantization may degrade the quality of the generated flow vectors~\cite{li2024mar}.

\textbf{Our solution.} To overcome these challenges, we reformulate the problem as a similarity matching task, where our primary objective is to project both flow representations and textual descriptions into a unified common space. Our approach seeks to achieve semantic matching while minimizing manual annotation and maintaining computational efficiency. To this end, we leverage the instruction-following data generated during the feature alignment stage.
We utilize a cross attention mechanism to automatically establish correspondences between textually described flow patterns and their corresponding streamline segments. By dynamically attending to relevant features across both modalities, the model learns to create meaningful associations between textual descriptions and flow representations. Additionally, we implement dedicated encoders for both text and streamline segments, ensuring their alignment within a shared embedding space and enhancing the effectiveness of the matching process.

\textbf{Model structure.}
Our semantic matching workflow is illustrated in~\cref{fig:matcher}.
We obtain the embeddings of streamline segments using our pre-trained \ours, while text embeddings are derived from SentenceBERT~\cite{reimers2019sentencebert}. We use SentenceBERT to make the semantic matching model more lightweight and efficient.
Furthermore, we employ a cross-modal attention mechanism~\cite{lee2018scan,li2022blip}, where the text embedding serves as the query, while each streamline segment embedding functions as both the key and value. This enables the model to effectively capture semantic correspondences between textual descriptions and flow pattern representations. By mapping both modalities into a shared common space through their respective modal encoders, we facilitate direct similarity computation and cross-modal interaction, ensuring alignment between them~\cite{betker2023dalle3}.

During the training phase, we use flow representations sampled from streamlines and process them through the \ours, mapping them into the common space. These embedded flow representations serve as the keys, while the text embeddings act as queries in the attention module.
The attention weights are computed to aggregate the streamline segment features. In this way, we obtain the attention scores that determine the similarity between the flow pattern and the streamline segments. The aggregated embeddings are then used for contrastive learning with the text embeddings. We employ the InfoNCE loss~\cite{oord2018infonceloss}, which optimizes the alignment between text and streamline segments, thereby enhancing cross-modal match performance. In the semantic matching phase, we use our \ours to encode all streamline segments and embed them into the common space. The semantic similarity is then computed between these embeddings and the semantic embedding of the flow pattern query. Since all sampled streamline segments can be pre-encoded, the matching process is efficient, enabling fast response in the visual interface.

\section{Interface}

\begin{figure*}[h]
    \centering
    \includegraphics[width=1\linewidth]{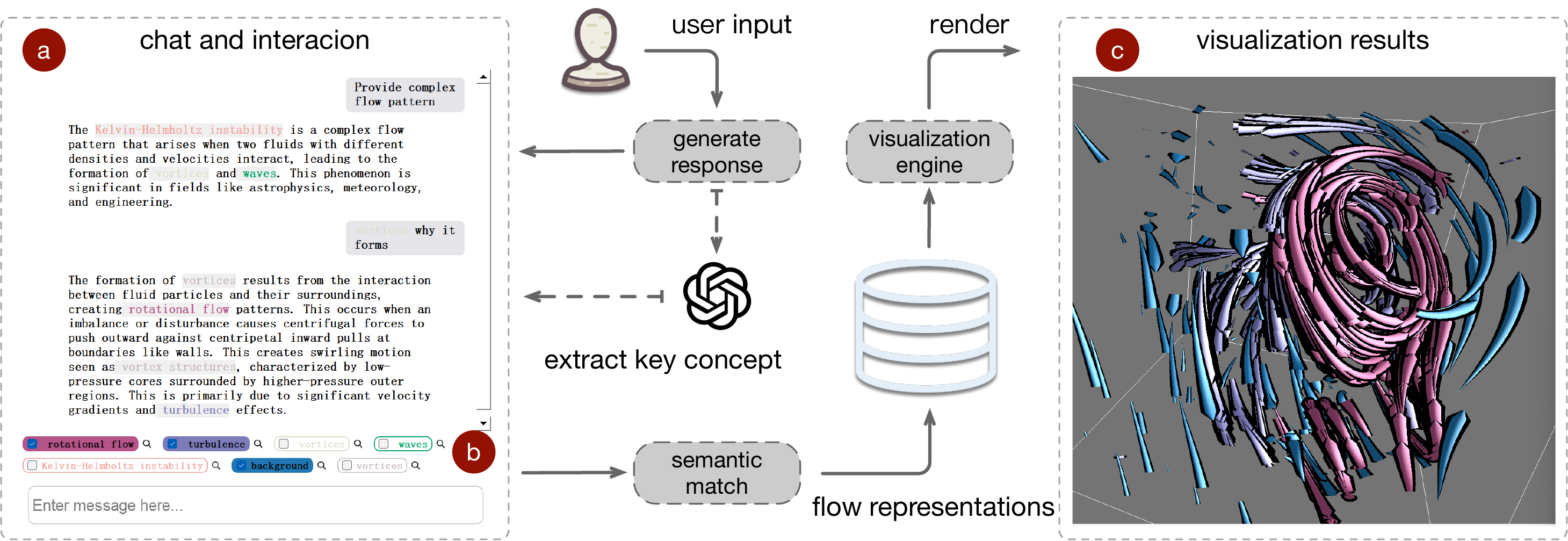}
    \caption{The interface illustrates an interaction workflow for exploring the flow field using natural language. (a) shows the chat view, enabling conversations with large language models. (b) displays an input box for entering dialogue content and query information. Colourful tags demonstrate the key concepts extracted in real time from the dialogue, allowing interactive clicking to highlight corresponding visualization elements. (c) presents the flow visualization results.}
    \label{fig:interface}
\end{figure*}

To apply our automated framework for exploring flow data better, we developed an interface that allows interaction through natural language and simple operations. As shown in~\cref{fig:interface}, our interface includes two main views: a dialogue window and a flow visualization window. The interactive workflow comprises \emph{chat}, \emph{key concept extraction}, and \emph{visualization}.

\comp{

\textbf{Chat with LLM.}
Our interface supports interactive exploration with flow data through natural language dialogue powered by an LLM. 
Our interface supports both natural language chat with LLM and multimodal querying by integrating flow representations. Since \ours encodes flow patterns based on the shape of streamline segments, the fine-tuned LLM learns to perceive structural similarities and generalize across diverse scenarios. The interface allows users to input text with flow pattern tags to trigger semantic queries. As shown in~\cref{fig:interface}~(a), selecting the ``vortices'' tag and querying ``why it forms'' leads the LLM to explain that vortices emerge due to velocity gradients that create a central vortex core and a surrounding high-pressure region. Such gradients result in characteristic flow features characterized by an outer high-pressure region and a central vortex core (shown in~\cref{fig:exploration}~(a)). The interface automatically maps selected flow structures into embeddings and combines them with tokenized instructions to enable seamless multimodal interaction.

\textbf{Extract key concepts.}
To simplify interaction, we employ GPT-4o as a real-time agent to extract flow-related key concepts from each LLM response. This helps users quickly identify important patterns without manual effort. Since multiple interaction rounds may produce redundant or overlapping tags, we implement a context management strategy that merges new tags into existing ones. All extracted concepts are displayed above the input box for easy access, as shown in~\cref{fig:interface}~(b).

\textbf{Visualize key concepts.}
To help users conveniently visualize flow structures of interest, our interface supports tag-based semantic matching. When a user clicks a tag, its textual description is encoded and matched against the pre-encoded flow representations to retrieve the most semantically relevant patterns. The interface then locates the corresponding streamline segments and renders them. 
Particle seeds are placed based on spatial positions to enable real-time dynamic tracing. 
As shown in~\cref{fig:interface}~(b) and (c), clicking ``turbulence'' and ``rotational flow'' tags retrieves and visualizes the corresponding structures with color-coded streamline segments.

}
\section{Evaluation}

In this section, we first evaluate the flow representations extracted by our proposed \ours. Specifically, we quantitatively assess their discriminability and uniformity to validate the effectiveness of our approach.
Next, we demonstrate how our prototype interface facilitates the intuitive exploration of complex flow fields. In particular, we highlight how natural language queries and concise interactive operations enhance users' efficiency in exploring, analyzing, and deriving insights from intricate flow fields.

\begin{figure}[h]
    \centering
    \includegraphics[width=1\linewidth]{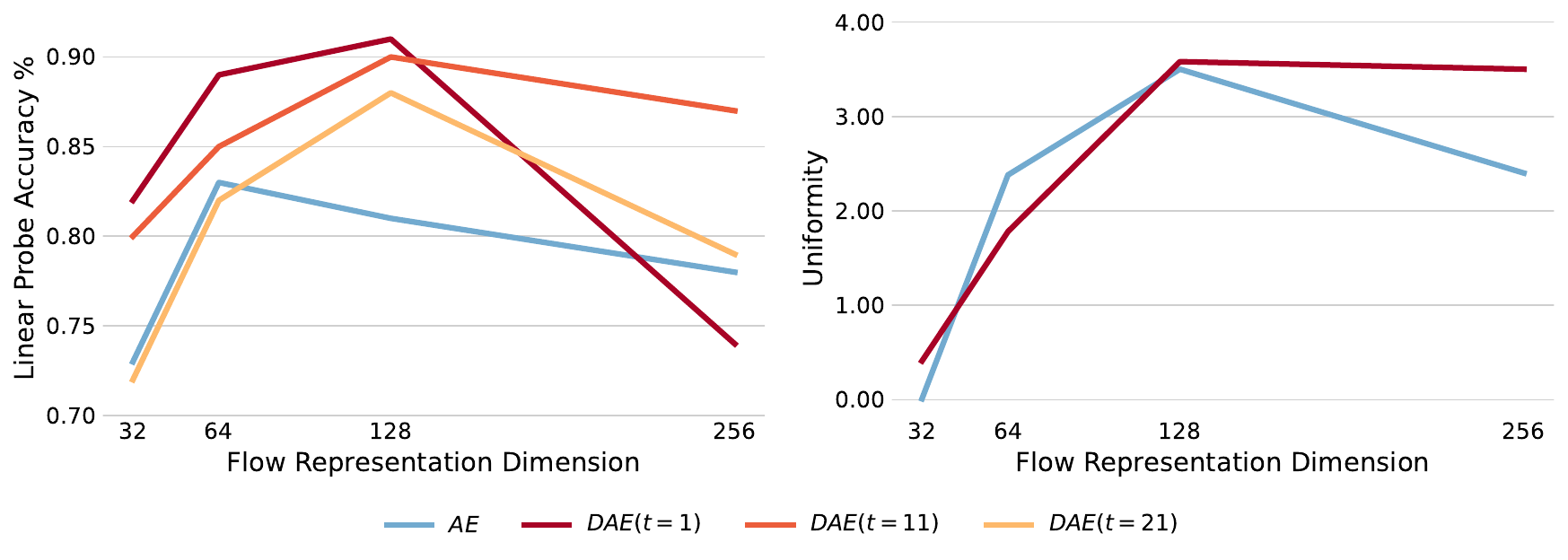}
    \caption{\hot{Feature discriminability (left) is influenced by the noise scale used during training and the dimensionality of the encoded flow features. Feature uniformity (right) reflects how different model structures and latent dimensions affect the distribution of the learned features.}}
    \label{fig:feature_extraction}
\end{figure}

\begin{figure*}
    \centering
    \includegraphics[width=1\linewidth]{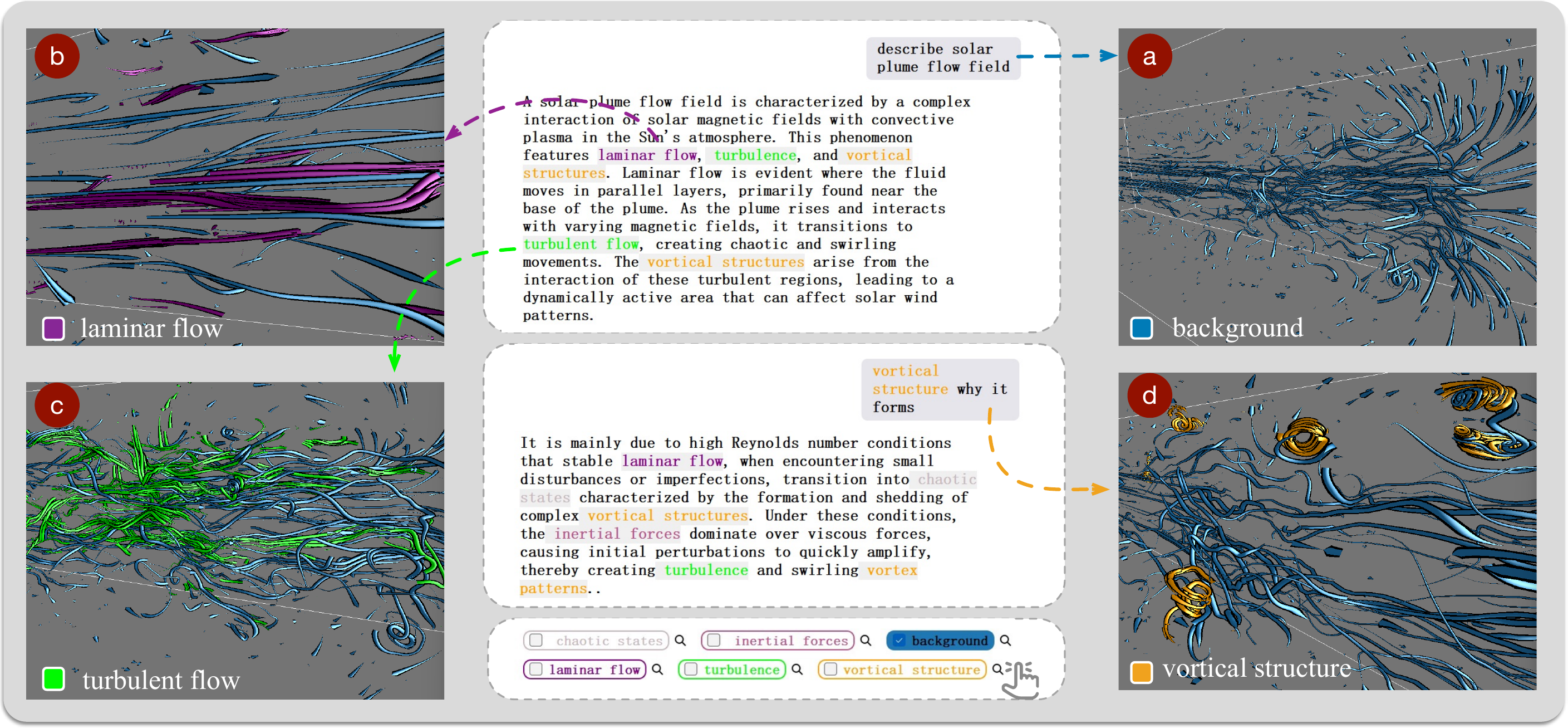}
    \caption{The primary visualization results are demonstrated through a case study using the solar plume dataset. The blue streamlets represent the background flow and are uniformly seeded throughout the entire domain.}
    \vspace{-.1in}
    \label{fig:case1}
\end{figure*}

\subsection{Quality of Flow Representation}

Since \ours is a pre-trained encoder designed specifically for streamline data, the quality of the extracted features directly impacts its performance in downstream tasks, such as flow pattern feature alignment, text semantic matching, and subsequent exploration. In this section, we evaluate the quality of the features extracted by \ours and investigate its capacity to distinguish among different flow patterns effectively.

\textbf{Extraction of discriminative features.}
The encoded features produced by \ours should possess sufficient discriminative capacity, allowing effective differentiation of flow patterns during the feature alignment stage. To assess this encoding capability, we employ a linear probe evaluation, which involves training a simple linear classifier using the extracted encoded features while freezing the pre-trained encoder weights. The performance of this classifier serves as an indicator of the discriminative quality captured by the encoder. Specifically, we freeze \ours and train a linear classifier to discriminate among various flow patterns. Thus, the classification accuracy reflects the discriminability and effectiveness of the extracted features.

As our objective is to validate features based on classification accuracy, we directly adopt as labels the datasets from which the streamlines originate. Notably, complex datasets inherently contain diverse flow patterns. Therefore, for the linear probe evaluation, we select only data subsets characterized by relatively uniform flow patterns. During pretraining, we directly leverage our pre-trained \ours to extract features from these streamline data. Specifically, we construct a dataset by uniformly sampling streamlines from four different flow datasets, namely Bernard, Cylinder, Five Critical Points, and Tornado. Consequently, the resulting dataset comprises four distinct classes corresponding to the source datasets, containing 2.4K streamlines.

Our training process includes two stages: training \ours and training the classification head. We first use eighty percent of the constructed dataset as the training set and pretrain \ours with different noise scales. Following~\cite{xiang2023ddae}, we select small noise scales at uniform intervals. After pretraining, we freeze \ours and train a simple classification head.
\hot{We report the average linear probe accuracy on the test set. As shown in~\cref{fig:feature_extraction}~left, the best performance occurs at an encoding dimension of 128, balancing compactness and expressiveness. Larger dimensions lead to degradation, possibly due to overfitting. The AE yields weaker results than DAEs, confirming the benefit of noise injection for learning discriminative features.}

\textbf{Extract uniform features.}
We use the pre-trained \ours to extract streamline segment features, aiming for these features to be evenly distributed across the feature space to avoid concentration. This ensures that when they are later mapped to the corresponding input space of the LLMs, a more comprehensive learning process can take place. Inspired by feature distribution analysis in contrastive learning~\cite{wang2020uniform}, we normalize the features to lie on the unit hypersphere and examine their uniformity properties. We define feature uniformity as a measure of the distribution uniformity on the unit hypersphere:

\begin{equation}
    \text{F}_{\text{uniform}} = - \log  \mathbb{E}_{x, y \sim \text{features}} \left[ e^{-2 \| f(x) - f(y) \|_2^2} \right].
\end{equation}

\noindent We examine the uniformity of features extracted by \ours and the autoencoder, with the best noise scale setting of \ours. As shown in~\cref{fig:feature_extraction}~right, it is difficult to obtain uniform features when the encoding dimension is low. The autoencoder structure also experiences a performance decline at higher feature dimensions. In summary, using the DAE method, our \ours is appropriately set with an encoding dimension of 128. 

In summary, we evaluate the quality of the flow representations produced by \ours from the perspectives of discriminability and uniformity, confirming their effectiveness for semantic alignment and matching tasks.

\subsection{Exploration Results}

In this section, we present case studies conducted on various flow datasets to illustrate the effectiveness of our approach to exploring flow patterns. Utilizing our interactive interface, users can highlight flow patterns of interest in real time. Meanwhile, the integrated LLM aids users by providing insightful explanations regarding the underlying mechanisms that contribute to these identified flow patterns.

\begin{figure*}
    \centering
    \includegraphics[width=1\linewidth]{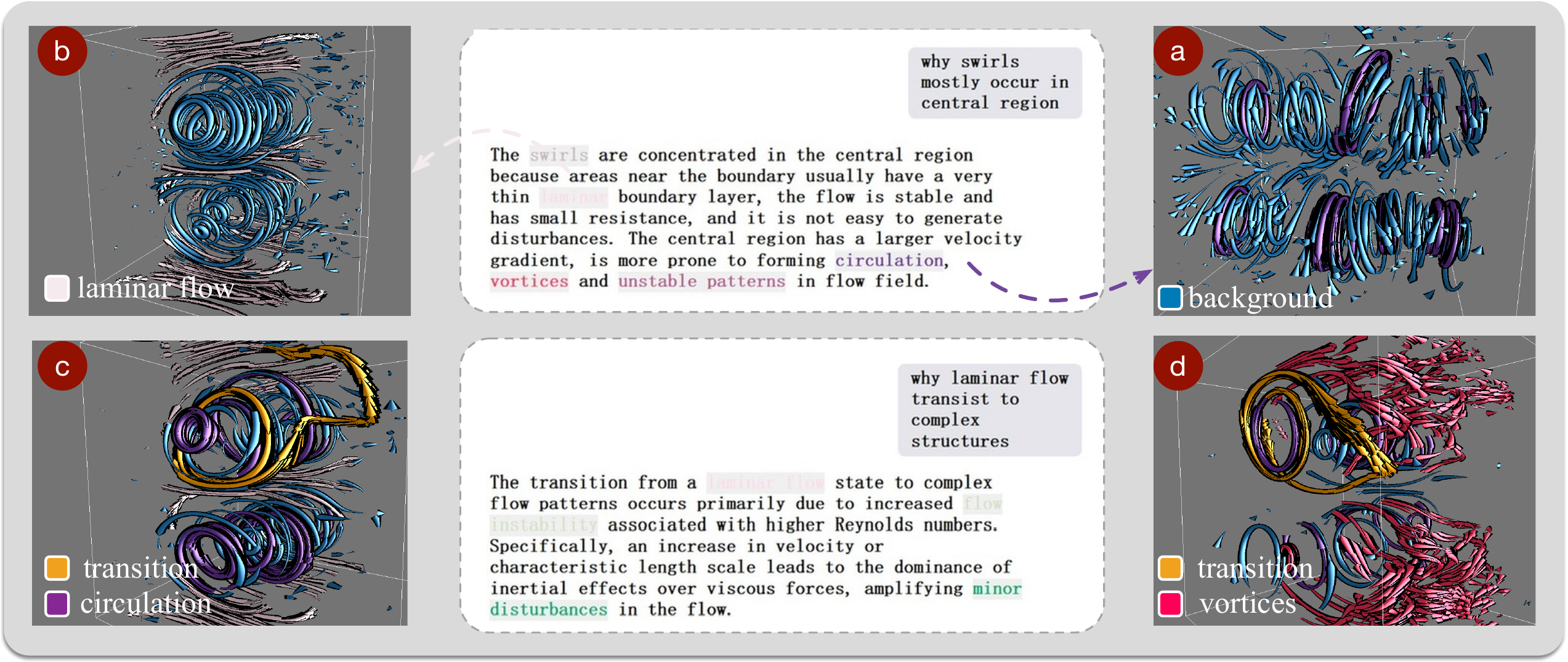}
    \caption{The primary visualization results are demonstrated through a case study using the two swirls dataset.}
    \label{fig:case2}
\end{figure*}

\comp{

\textbf{Case study on solar plume dataset.}
We first visualized the global flow field using randomly seeded particles, as shown in~\cref{fig:case1}~(a). This visualization provides an intuitive representation of the global flow structure (\tikz\draw[thick, fill=background] (0,0) rectangle (0.18,0.18);). The left region exhibits smooth, laminar flow, while the right shows increased curvature and distortion, suggesting turbulence. To validate this observation, we queried the LLM, which described the flow as a ``complex interaction of solar magnetic fields.''
Next, we asked ``What causes complex turbulence?'' The LLM identified two key factors: high Reynolds number, triggering transition to turbulence, and strong velocity gradients, which induce vortical structures through shear interactions. The resulting turbulence visualization (\tikz\draw[thick, fill=green] (0,0) rectangle (0.18,0.18);) in~\cref{fig:case1}~(c) confirms these insights, with chaotic regions densely populated by intertwined streamlets.

Our interface uses an LLM-based agent to extract flow pattern tags from generated text. In the turbulence explanation, laminar flow and vortical structures were identified~\cref{fig:case1}. Selecting the laminar flow tag displays its visualization~\cref{fig:case1}~(b), showing that such flows (\tikz\draw[thick, fill=purple] (0,0) rectangle (0.18,0.18);) are concentrated on the left side. They behave with smooth and uniform motion.
Next, selecting the vortical structures tag (\tikz\draw[thick, fill=orangepeel] (0,0) rectangle (0.18,0.18);) highlights spiral-like flow patterns scattered throughout the field~\cref{fig:case1}~(d). The clustered vortices clearly indicate intensive rotation and instability in those fluid regions. The surrounding blue streamlines likely engage with and are influenced by these rotational movements, suggesting relatively localized formation areas.
To investigate the cause, we clicked the query button for this tag and asked ``why it forms?'' The LLM responded that small disturbances in laminar flow, particularly under high Reynolds number conditions, can trigger vortex formation. This transition from laminar to complex flow patterns is a common characteristic observed in various scientific fields. For example, air masses can initiate circulatory motions due to thermal expansion and contraction effects.

\textbf{Case study on two swirls dataset. }
From~\cref{fig:case2}~(a), we can first observe the overall flow pattern across the entire flow field. The core region of the flow field exhibits large-scale swirl structures, indicating dominant rotational motion. In~\cref{fig:case2}~(a), we visualize the circulation structures (\tikz\draw[thick, fill=purpleheart] (0,0) rectangle (0.18,0.18);) embedded within the core region. These structures form ring-like shapes due to strong rotation, while the surrounding regions show weaker and more scattered patterns. We analyzed this phenomenon using the LLM. One possible explanation is that the spatial constraints in the boundary region lead to an increase in flow velocity, which in turn enhances instabilities.
In the edge regions of the flow, the boundary layer is typically thin and exhibits a more laminar behaviour. We clicked on the laminar flow tag to inspect the laminar flow (\tikz\draw[thick, fill=piggypink] (0,0) rectangle (0.18,0.18);) in the edge regions. The visualization results are shown in~\cref{fig:case2}~(b). As a result, the laminar flow remains relatively organized compared to the more dynamic structures observed in the central regions. 

We were interested in how these flow patterns transition between different states. To explore this, we visualized the laminar flow, circulation structures, and transitional paths (\tikz\draw[thick, fill=orangepeel] (0,0) rectangle (0.18,0.18);) as shown in~\cref{fig:case2}~(c). We observed that the flow moves from the edge to the center along a complex and winding trajectory. This suggests that the transition is not linear, but involves intricate redirections and structural changes, likely driven by local velocity gradients and vorticity interactions. Similar transitions were also observed in the solar plume case. According to LLM-generated insights, this transition may be attributed to an increase in velocity or characteristic length scale, leading to a higher Reynolds number. 
The explanation from the LLM, ``due to higher Reynolds numbers,'' is demonstrated in~\cref{fig:case2}.
As the Reynolds number increases, the flow becomes more unstable, ultimately triggering the transition from laminar to turbulent or vortical structures. 

To further investigate the patterns along the flow transition pathways, we examined additional flow structures. As shown in~\cref{fig:case2}~(d), we present vortex structures (\tikz\draw[thick, fill=radicalred] (0,0) rectangle (0.18,0.18);) with differing flow directions. These vortices exhibit distinct rotational behaviours, suggesting the presence of complex interactions between flow regions. Such variations in vortex orientation may indicate shear effects, velocity gradients, or local instabilities that influence the overall transition process. We could still visualize the transition paths (\tikz\draw[thick, fill=orangepeel] (0,0) rectangle (0.18,0.18);) from these structures to circulation flow, as illustrated in~\cref{fig:case2}~(d). During this transition, the flow undergoes a direct and relatively straightforward path change over a short distance, moving from the vortical structure straight into the central area and gradually transforming into circulation flow. In analyzing the transition process changes using LLM queries, it was observed that two structures with opposing directions can produce strong shear forces due to differential flow velocities. These shear forces may cause the vortices to be skewed during the transition flow, leading to significant instability. 

\textbf{Explore flow structures.}
Additionally, we further explored key flow structures using two additional datasets to validate the effectiveness of our framework. As shown in~\cref{fig:exploration}~(a), although the five critical points dataset exhibits relatively simple overall flow variations, it reveals subtle yet significant small-scale vortex structures. These localized eddies arise from delicate velocity gradients and boundary-layer interactions.
Similarly, in~\cref{fig:exploration}~(b), the cylinder dataset displayed predominantly smooth background streamlines, indicating laminar flow, while a distinct localized eddy emerged in the central region. This structure exhibited rotational characteristics that introduced disturbances and fluctuations into the flow field. These variations may serve as precursors to instabilities or transitions to complex flow behaviors.
}

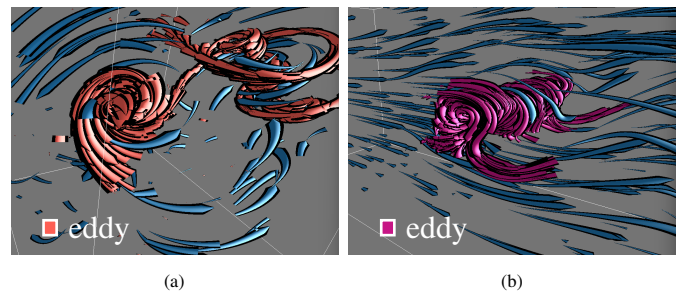
\begin{figure}
    \centering
    \begin{minipage}{0.24\textwidth}
        \centering
        \begin{tikzpicture}
            \definecolor{customcolor}{HTML}{fd5e53}
            \node[anchor=south west,inner sep=0] (img) at (0,0) {\includegraphics[width=\textwidth]{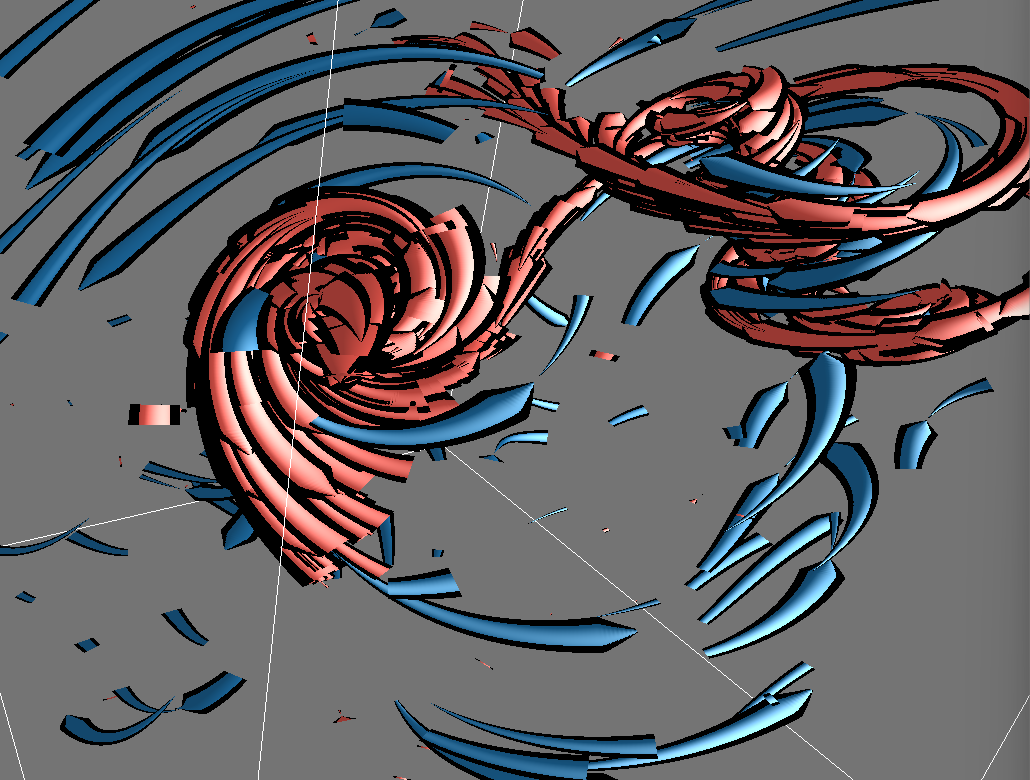}};
            \begin{scope}[x={(img.south east)},y={(img.north west)}]
                \draw[fill=customcolor, draw=white, line width=1pt] (0.1,0.08) rectangle (0.14,0.14);
                \node[anchor=west, text=white, font=\Large] at (0.15,0.10) {eddy};
            \end{scope}
        \end{tikzpicture}
        \subcaption{} 
        \label{fig:subfig_a} 
    \end{minipage}
    \hfill
    \begin{minipage}{0.24\textwidth}
        \centering
        \begin{tikzpicture}
            \definecolor{customcolor}{HTML}{c71585}
            \node[anchor=south west,inner sep=0] (img) at (0,0) {\includegraphics[width=\textwidth]{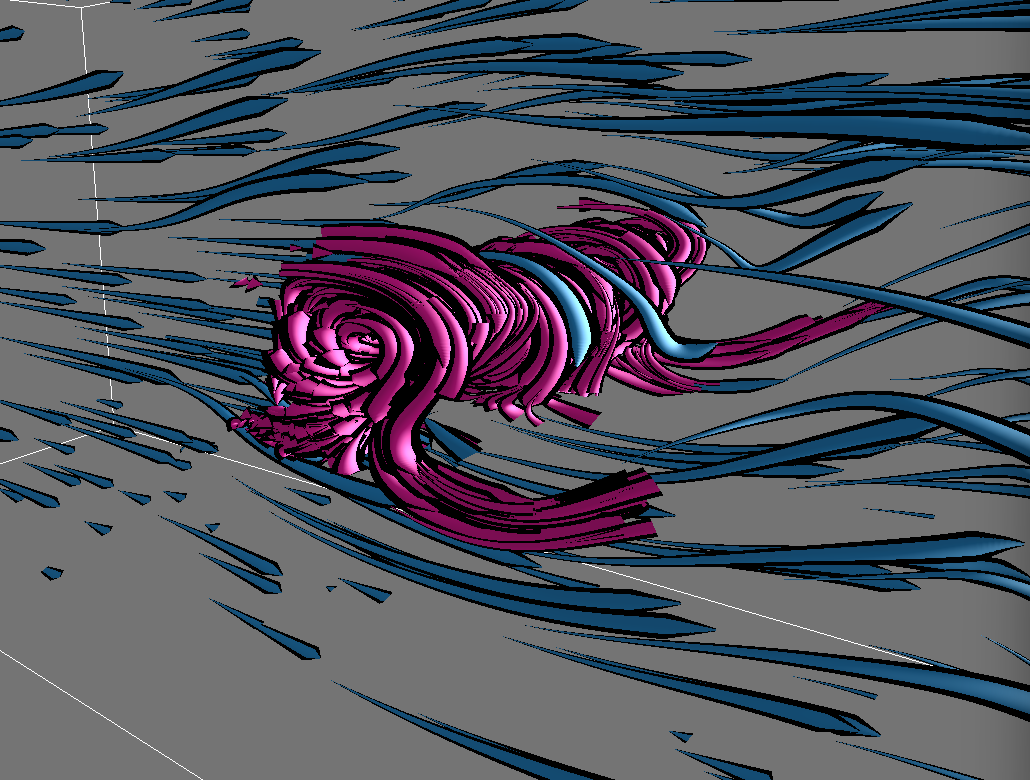}};
            \begin{scope}[x={(img.south east)},y={(img.north west)}]
                \draw[fill=customcolor, draw=white, line width=1pt] (0.1,0.08) rectangle (0.14,0.14);
                \node[anchor=west, text=white, font=\Large] at (0.15,0.10) {eddy};
            \end{scope}
        \end{tikzpicture}
        \subcaption{} 
        \label{fig:subfig_b} 
    \end{minipage}
    
    \caption{Flow visualization results using additional datasets are presented to further analyze flow structures. (a) illustrates the results obtained from the five critical points dataset, while (b) presents the visualization using the cylinder dataset.}
    \label{fig:exploration}
\end{figure}

\subsection{Case Study by Domain Experts}

\begin{figure*}
    \centering
    \includegraphics[width=1\linewidth]{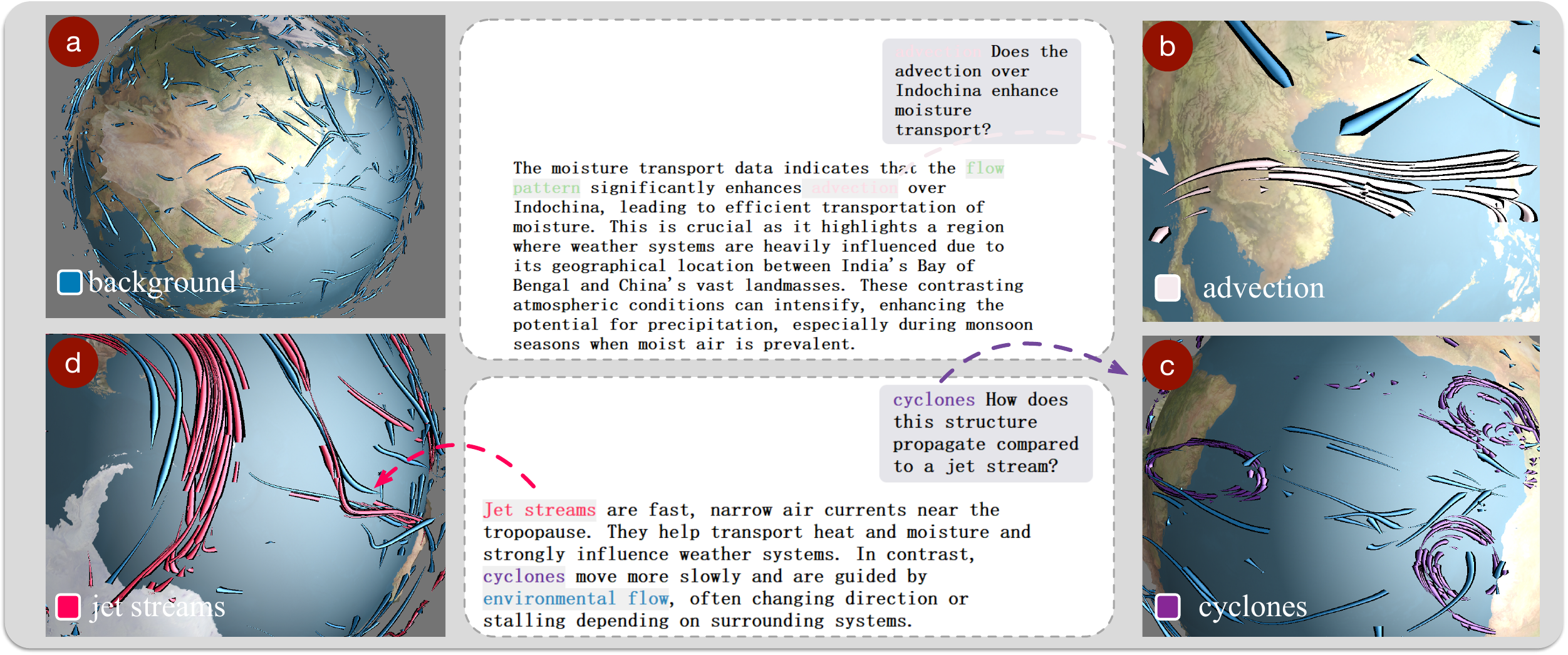}
    \caption{\hot{The key visualization results from exploration with domain experts using the ECMWF data.}}
    \label{fig:case3}
\end{figure*}

\hot{

\textbf{Expert.}
We invited three experts ($E1$–$E3$) to participate in our case study, each with different backgrounds and research expertise. $E1$ is a scientist specializing in physics and atmospheric science, with extensive experience in analyzing flow phenomena. $E2$ and $E3$ are senior PhD students with up to four years of research experience, focusing on meteorology and hydrology.

\textbf{Data and overall pattern.}
The experts were tasked with exploring real-world atmospheric data. As part of the case study, we incorporated data from the European Centre for Medium-Range Weather Forecasts (ECMWF), a dataset recognized by $E3$ as the most authoritative source in the meteorological community. As shown in~\cref{fig:case3}~(a), we visualized the global atmospheric flow field (\tikz\draw[thick, fill=background] (0,0) rectangle (0.18,0.18);), where streamlets represent wind trajectories at a specific altitude and time.

\textbf{Exploration.} 
The experts were interested in moisture transport and began their analysis by examining advection flow. The visualization identified advection pattern (\tikz\draw[thick, fill=piggypink] (0,0) rectangle (0.18,0.18);) over Southeast Asia, which reflects the large-scale transport of air masses by the wind field, as shown in \cref{fig:case3}~(b). 
The streamlets exhibit strong directional coherence, indicating organized airflow that likely contributes to regional moisture transport. Motivated by this observation, $E1$ and $E2$ prompted the LLM with a follow-up question about the role of advection over Indochina, aiming to assess whether such flow structures enhance regional moisture transport. They asked: ``Does the advection over Indochina enhance moisture transport?'' In response, the LLM identified the flow pattern as a significant contributor to enhanced advection across Indochina, thereby facilitating efficient moisture transport.
$E2$ and $E3$ confirmed that Indochina, situated between the Bay of Bengal and China, is a convergence zone for contrasting atmospheric systems. They noted that during the monsoon, strong moisture transport enhances precipitation. $E3$ further highlighted moisture transport as a key example of vector field analysis in atmospheric research.

$E1$ wanted to explore more dynamic and complex atmospheric phenomena. He used the interface to visualize cyclone-related flow structures (\tikz\draw[thick, fill=purpleheart] (0,0) rectangle (0.18,0.18);), as shown in~\cref{fig:case3}~(c). The visualization revealed several cyclone structures over the tropical Atlantic region. These flow patterns exhibit prominent spiral formations and rotational symmetry, consistent with the typical characteristics of tropical cyclones. $E1$ further discussed cyclone duration with the LLM and was satisfied with the response, which aligned with operational forecasting data.

The experts further investigated the difference in cyclones and jet streams propagation. $E2$ queried the LLM: ``How does this structure propagate compared to a jet stream?'' As shown in~\cref{fig:case3}, the LLM distinguished between the two atmospheric structures and provided respective definitions.
$E2$ and $E3$ affirmed the explanation and emphasized the distinctions in meteorological forecasting, where jet streams frequently act as steering mechanisms that modulate the trajectory and evolution of cyclonic systems.
They also prompted the system to visualize the jet streams (\tikz\draw[thick, fill=radicalred] (0,0) rectangle (0.18,0.18);), as shown in~\cref{fig:case3}~(d). They found this to be the Antarctic Jet, extending from the eastern coast of South America toward the waters west of Antarctica, which transports moisture and heat across latitudes. The experts also noted its potential link to large-scale atmospheric phenomena, such as Rossby wave activity. 

\textbf{Discussion.}
While our case studies demonstrate the effectiveness of semantic alignment for flow pattern exploration, incorporating scalar information could further enhance the analysis of atmospheric phenomena. For instance, as noted by $E3$, identifying Rossby waves could be more accurate by observing periodic variations in pressure fields, which resemble wave-like structures. Exploration guided by both flow patterns and scalar attributes has been previously supported in FlowNL~\cite{huang2022flownl}, and FlowLLM~\cite{li2025flowllm} further integrated LLMs for such composite queries. However, as this work focuses primarily on automatically aligning semantic representations with flow patterns, we do not include scalar information in the interaction process.
}
\section{Discussion and Conclusion}

\textbf{Limitations.}
\hot{While our approach is applicable to any spatial curve with appropriate sampling, its effectiveness for unsteady fields remains to be evaluated. Pathlines may exhibit more complex patterns and distinct mappings to the text embedding space, potentially requiring new training data, model retraining, and even additional modules to address unexpected failure cases. It is important to note that this approach facilitates the exploration of the shape of integral curves using natural language. As a result, it cannot address flow phenomena that are not well represented by curve geometry alone.}

Additionally, the use of sampled streamlines may miss subtle structures or result in an imbalanced representation across flow patterns, particularly for rare but important cases. The density of the samples along the streamline may also affect the resulting patterns. While rendered images help compensate through instruction-following data generation, the lack of annotated scientific datasets still limits generalizability. Moreover, the current representation emphasizes streamline shape and omits other flow characteristics, though the alignment framework is flexible and can be extended to incorporate richer descriptors.

\textbf{Conclusion.}
In this paper, we present a semantic alignment framework for exploring flow visualizations through natural language interaction. By converting streamline segments into distance matrices and aligning them with an LLM’s semantic space, our method enables rigid-invariant flow representations. 
These representations, along with rendered images and LLM-generated descriptions, create a text-flow instruction-following training dataset for the LLM. Fine-tuning the LLM on this dataset builds cross-modal attention between flow representations and the semantics behind images and text. This mechanism allows for effective matching between flow structures and textual queries. \hot{Our interactive interface demonstrates the power of natural language-driven exploration, facilitating accessible analysis of flow phenomena for a broader audience.}

In the future, we would like to utilize LLMs further as intelligent agents to handle natural language requests more automatically and efficiently. This includes not only interpreting user intentions more precisely and quickly, but also employing more advanced and effective methodologies within the semantic matching module. By continuously optimizing and improving these semantic matching techniques, the overall workflow for dealing with requests can become robust and accurate, enhancing user engagement and system usability.

\section*{Acknowledgements}
This research was supported in part by the National Natural Science Foundation of China through grants 62372484 and 62172456.
The authors would like to thank the anonymous reviewers for their insightful comments.


\bibliographystyle{abbrv-doi-hyperref}
\bibliography{template}

\clearpage

\appendices
 
\setcounter{page}{1}
\setcounter{figure}{0}
\setcounter{section}{0}
\setcounter{table}{0}

\hot{
\section{Pretraining flow encoder}

To pretrain the \ours, we first compute the distance matrices for the collected streamline segments. These distance matrices serve as shape descriptors of the streamline segments, which are invariant under rigid transformation. Based on this representation, we adopt a self-supervised learning approach to derive latent representations of flow patterns.
Specifically, we employ a denoising autoencoder to encode the streamline distance matrices into compact flow pattern representations. This process constitutes the pretraining phase of the flow encoder, enabling it to learn structural patterns without explicit labels.

\textbf{Data preparation.}
We use an atmospheric dataset, BOMEX, to measure the computation time and training quality in this experiment. This dataset simulates shallow cumulus convection in a domain of 3175 × 3175 × 3980 meters, including one flow field and fourteen scalar fields. From the flow field, 30,000 streamline segments are randomly sampled, which are then used to pretrain our \ours. 
All computations and visualizations were performed on a machine equipped with an NVIDIA RTX 4090 graphics card.
First, the distance matrices are computed and normalized in parallel using CUDA. This step takes 2.62 seconds for all segments. 
Each distance matrix represents the flow pattern of a streamline segment. The matrices are used to sufficiently train our \ours until the distance matrices can be reconstructed from the latent vectors. The pre-trained \ours is used to encode all the streamline segments into flow pattern representations.

\textbf{Model design.}
Our \ours is similar to typical UNet for images, consisting of blocks that blend linear projection (i.e., convolution and linear layers) with non-linear activations. The \ours transforms a distance matrix into a latent vector, and reconstructs the distance matrix from the vector. When the distance matrix can be accurately reconstructed, the vector is considered to contain the same information as the distance matrix. To enhance the unsupervised learning process, we follow the noise schedule used in the denoising diffusion probabilistic model (DDPM) to add Gaussian noise to the distance matrices. The denoising autoencoder is then employed as \ours to reconstruct the original matrices, thereby learning flow representations. 

Our \ours is based on a UNet architecture with both encoder and decoder paths composed of stacked convolutional residual blocks. The network processes 2D inputs of shape 1×32×32 through multiple resolution levels, with downsampling performed via strided convolutions and upsampling via transposed convolutions. At each level, temporal information is injected using sinusoidal positional encoding projected by linear layers and added to the feature maps. 
The encoded latent representation is further compressed via a fully connected layer to a fixed latent size of 128 and subsequently decoded using a symmetric fully-connected decoder.
}

\begin{table}[!thbp]
  \caption{The average RMSE, PSNR, and SSIM with different encoding approaches on the BOMEX dataset.}
  
  \scriptsize%
  \centering%
  \setlength{\tabcolsep}{1.6mm}{
  \begin{tabular}{%
            c|%
  	  	*{6}{r}%
  	}
  	\toprule
  	Approach & RMSE $\downarrow$ & PSNR $\uparrow$ & SSIM $\uparrow$ \\
  	\midrule
        AE &  0.0061 & 45.13 & 0.99 \\
        DAE &  \textbf{0.0037} & \textbf{48.95} & 0.99 \\
                 
  	\midrule               
  	\bottomrule
  \end{tabular}}
  \label{tab:recon}
\end{table}

\begin{table}[!thbp]
  \caption{Comparison of training efficiency and convergence accuracy across different methods on the BOMEX dataset. All models were trained for 100 epochs. The training time is reported in seconds, and MSE loss is reported in units of $10^{-5}$.}
  
  \scriptsize%
  \centering%
  \setlength{\tabcolsep}{1.6mm}{
  \begin{tabular}{%
            c|%
  	  	*{6}{r}%
  	}
  	\toprule
  	Approach & Timing $\downarrow$ & Loss $\downarrow$\\
  	\midrule
        AE &  \textbf{204.85} & 5.78 \\
        DAE &  210.82 & \textbf{2.00} \\
                 
  	\midrule               
  	\bottomrule
  \end{tabular}}
  \label{tab:timing}
\end{table}

\begin{figure}[ht]
    \centering
    \begin{subfigure}[b]{0.11\textwidth}
        \includegraphics[width=\textwidth]{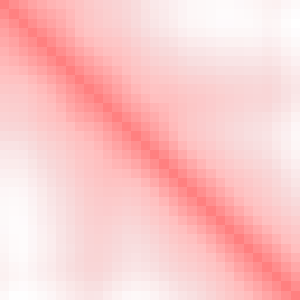}
    \end{subfigure}
    \begin{subfigure}[b]{0.11\textwidth}
        \includegraphics[width=\textwidth]{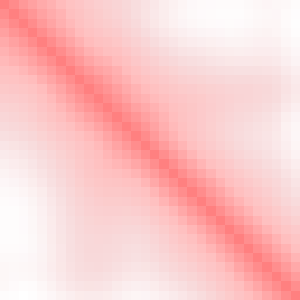}
    \end{subfigure}
    \begin{subfigure}[b]{0.11\textwidth}
        \includegraphics[width=\textwidth]{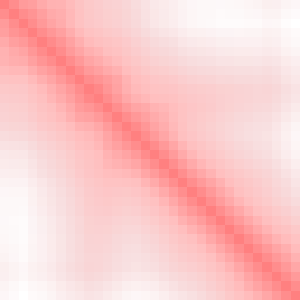}
    \end{subfigure}
    \begin{subfigure}[b]{0.11\textwidth}
        \includegraphics[width=\textwidth]{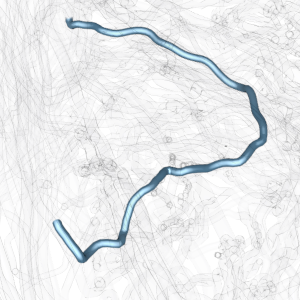}
    \end{subfigure}

    \vskip\baselineskip 

    \begin{subfigure}[b]{0.11\textwidth}
        \includegraphics[width=\textwidth]{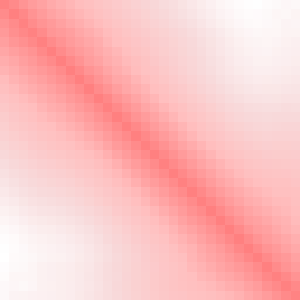}
        \caption{AE}
    \end{subfigure}
    \begin{subfigure}[b]{0.11\textwidth}
        \includegraphics[width=\textwidth]{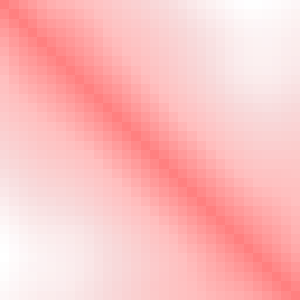}
        \caption{DAE}
    \end{subfigure}
    \begin{subfigure}[b]{0.11\textwidth}
        \includegraphics[width=\textwidth]{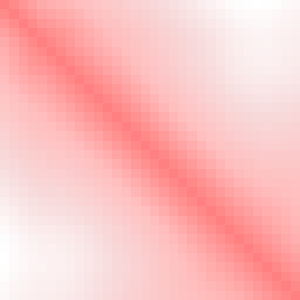}
        \caption{GT}
    \end{subfigure}
    \begin{subfigure}[b]{0.11\textwidth}
        \includegraphics[width=\textwidth]{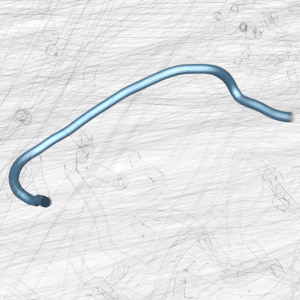}
        \caption{Segment}
    \end{subfigure}

    \caption{A qualitative comparison of the reconstructed distance matrices, using different approaches. (a) and (b) show the reconstructed results using AE and DAE, respectively. (c) shows the original distance matrices as the ground truth, and (e) shows the corresponding streamline segments.}
    \label{fig:recon}
\end{figure}

\hot{\textbf{Implementation, training, and performance.}
We utilized the Adam optimizer for parameter updates. The batch size was set to 128 distance matrices. The learning rate was set to $10^{-3}$. The model was trained on the BOMEX dataset for 100 epochs. The training results, including the final MSE loss, are summarized in~\cref{tab:timing}. 

We examine the performance of our \ours and compare it with a typical autoencoder (AE). As shown in~\cref{tab:recon}, our \ours based on DAE achieves the smallest reconstruction loss (RSME) and the highest peak signal-to-noise ratio (PSNR) and structural similarity index (SSIM). 
We further study the reconstruction performance by examining their visualization results. The reconstructed distance matrices of two example streamline segments are shown in~\cref{fig:recon}.
In our study, we find that both the DAE and the AE perform well in reconstructing the original distance matrix. Considering the comparable training costs and the superior outcomes achieved with DAE, we selected it as the architecture for the flow encoder.
}

\begin{figure}[h]
    \centering
    \includegraphics[width=1\linewidth]{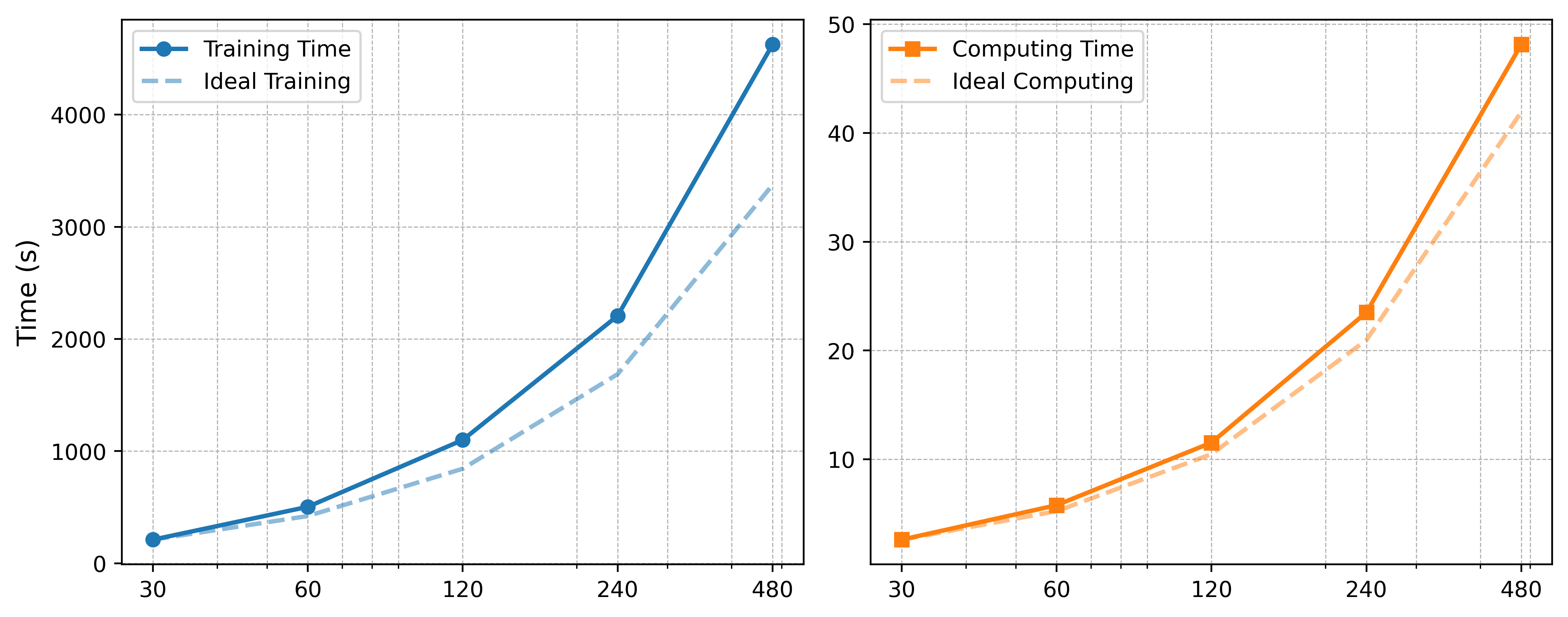}
    \caption{Timing analysis on different amounts of streamline segments (in thousands). The left plot shows the training time (in seconds) of the DAE, while the right plot reports the time for distance matrix computation. Dashed lines indicate ideal linear scalability for reference.}
    \label{fig:scaling}
\end{figure}

\hot{
We further examine the scalability by measuring both the distance matrix computation time and the training time of the \ours across varying numbers of streamline segments. As shown in~\cref{fig:scaling}, the left plot shows the total training time, while the right plot shows the time spent computing the distance matrices. We continue to use the BOMEX dataset. The horizontal axis is measured in thousands. The \ours is trained for 100 epochs with a batch size of 128.
The results indicate that the distance matrix computation exhibits near-linear scalability, with only a moderate increase in processing time as the dataset size grows. In contrast, the training time increases more rapidly, particularly when the number of streamline segments exceeds 240,000, reflecting the higher computational complexity involved in neural network optimization. Nevertheless, the overall training time remains manageable, under 4500 seconds for the largest dataset.
}

\hot{
\section{Align to large language model}
To enable the Large Language Model (LLM) to better understand the contextual relationships between different types of tokens, we introduce a projector network that maps the flow representation inputs into the backbone space of the LLM. This integration allows the model to generate responses conditioned on both textual and flow-based inputs. 

\textbf{Projection layers.}
While aligning tokens from different modalities to the text space using projection layers has proven effective and is widely adopted across various domains~\cite{liu2024llava,xu2024pointllm,zeng2024mllm4chartQA}, the appropriate number of projection layers requires discussion.
We investigate this by experimenting with one to four projection layers using varying hidden dimensions. The results of these configurations are presented in~\cref{tab:projection}.
}

\begin{table}[!thbp]
  \caption{The loss values with different numbers of projection layers using the solar plume dataset. The column “Hidden Dims” specifies the number of dimensions in the intermediate layers.}
  
  \scriptsize%
  \centering%
  \setlength{\tabcolsep}{1.6mm}{
  \begin{tabular}{%
            c|%
  	  	*{6}{r}%
  	}
  	\toprule
  	Hidden Dims & Loss $\downarrow$ \\
  	\midrule
        N.A. &  0.57 \\
        1024 &   0.46  \\
        1024, 2048 &   0.58  \\
        1024, 2048, 4096 &   \textbf{0.41}  \\
                 
  	\midrule               
  	\bottomrule
  \end{tabular}}
  \label{tab:projection}
\end{table}

\hot{
These results suggest that increasing the number and depth of projection layers can improve model performance, as evidenced by the lowest loss (0.41) obtained with three layers of increasing hidden dimensions. However, the performance drop observed with two intermediate layers indicates that more layers do not always guarantee better results, highlighting the importance of careful architectural design.

\textbf{Implementation and training.}
Based on these findings, we adopt a multi-layer perceptron network with three hidden layers as the projector to inject multimodal information into the LLM backbone.
For the backbone, we use the 8B version of LLaMA3~\cite{dubey2024llama3}, in order to avoid potential issues such as hallucination that are more common in larger-scale models~\cite{xu2024pointllm}.

During training, we set the batch size to four and train the model for a total of two epochs. We use the AdamW optimizer to train the model, applying it only to parameters needing to be updated. The learning rate is set to $1\times10^{-4}$, and a weight decay of 0.01 is applied to prevent overfitting and encourage better generalization.
To enable efficient fine-tuning of the LLM backbone, we apply LoRA with the following configuration: rank = 16, scaling factor = 32, dropout rate = 0.05, and no bias tuning. The adaptation is applied to the query and value projection layers of the attention mechanism. The LoRA-enhanced model is wrapped and moved to the GPU for training.
We adopt a two-stage training strategy: the first stage focuses on feature alignment by training the projector only, while the second stage performs instruction tuning by jointly optimizing the projector and the LLM. These results are summarized in~\cref{tab:training}, highlighting the effectiveness of each training stage and the impact of LoRA fine-tuning.
}

\begin{table}[!thbp]
  \caption{Comparison of training efficiency and loss value across different training strategies on the Plume dataset. All models were trained for two epochs; training time is reported in hours.}
  
  \scriptsize%
  \centering%
  \setlength{\tabcolsep}{1.6mm}{
  \begin{tabular}{%
            c|%
  	  	*{6}{r}%
  	}
  	\toprule
  	Training & Timing  & Loss $\downarrow$\\
  	\midrule
        Feature alignment &  1.24 & 1.71 \\
        Instruction tuning &  1.04 & \textbf{0.41} \\
        w/o LoRA &   2.79   & 1.58 \\
                 
  	\midrule               
  	\bottomrule
  \end{tabular}}
  \label{tab:training}
\end{table}

\hot{
As shown in~\cref{tab:training}, the two-stage training strategy significantly improves the model's performance. The instruction tuning stage achieves the lowest loss value of 0.41, indicating that joint optimization of the projector and LLM effectively enhances the model's text generation capability.
Moreover, the results highlight the essential role of LoRA in our training setup. Without LoRA-based fine-tuning, the model exhibits a higher loss (1.58), suggesting less efficient adaptation to multimodal inputs. This demonstrates that LoRA enables parameter-efficient fine-tuning while preserving model expressiveness, making it particularly well-suited for tasks involving cross-modal integration.

\textbf{Evaluation.}
We further evaluate the effectiveness of our semantic alignment framework by comparing it against a baseline where the LLM is trained without LoRA fine-tuning. Specifically, we sample 100 instances from the text-flow dataset and prompt the LLM to generate responses. These responses are then evaluated using the more powerful GPT-4o as an external judge to assess response quality.
}

\begin{figure}
    \centering
    \includegraphics[width=1\linewidth]{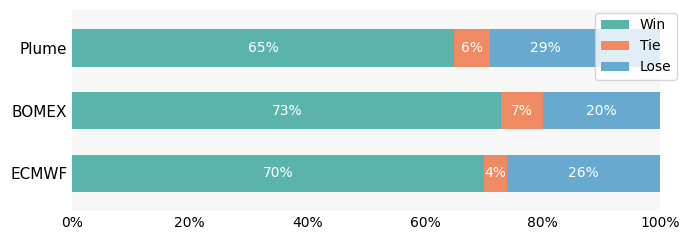}
    \caption{Comparison results between our method and the baseline without LoRA fine-tuning on three datasets, evaluated by GPT-4o. Each bar shows the percentage of responses judged as Win, Tie, or Lose.}
    \label{fig:win_rate}
\end{figure}

\hot{
As illustrated in~\cref{fig:win_rate}, our method consistently outperforms the baseline across all three datasets. Specifically, it achieves $73\%$ wins on BOMEX, $70\%$ on ECMWF, and $65\%$ on solar plume, indicating a strong advantage in generating higher-quality responses. The relatively low percentages of lost cases ($20\%$–$29\%$) further confirm the superiority of the LoRA fine-tuned model. The tie rate remains low across datasets, suggesting that GPT-4o could clearly distinguish between the outputs of the two training methods. These results demonstrate the effectiveness of our semantic alignment framework and the critical role of LoRA in enhancing response quality in multimodal scenarios.

To further evaluate the effectiveness of our semantic alignment framework, we conduct a qualitative comparison between our model and a baseline trained without LoRA fine-tuning. 
As shown in~\cref{tab:plume_response}, our method produces more semantically rich and physically accurate responses compared to the baseline without LoRA fine-tuning. When asked to describe the shape of the streamline, our model identifies spiral-like patterns, reflecting a deeper understanding of the curvature complexity. Similarly, for the prompt to illustrate the streamline, it highlights swirling, turbulent vortices and attributes them to shear stress and velocity gradients, demonstrating domain-relevant reasoning.
In contrast, the baseline model provides more generic descriptions, often missing the intricate flow characteristics visible in the images. These results confirm that LoRA fine-tuning improves the model's ability to align multimodal inputs and generate contextually appropriate textual explanations.
}

\begin{table*}
\renewcommand{\arraystretch}{2} 
\centering
\caption{Qualitative comparison of text responses generated by our model and the baseline model without LoRA given description prompts. 
}
\label{tab:plume_response}
\begin{tabular}{p{3.5cm} p{5.5cm} p{5.5cm} c}
\toprule
\textbf{Prompt} & \textbf{Ours} & \textbf{w/o LoRA} & \textbf{Image} \\
\midrule
Describe the shape of the streamline.  
& The streamline shows relative steadiness, further suggesting that the motion follows an outward curved shape. As it turns inward, the curvature indicates a more complex nature, signifying that the streamline \textbf{exhibits spiral-like patterns}.   
& The flow pattern displays a smooth, undulating structure, starting with an upward curve and progressing in gentle arcs. Several interconnected loops form, maintaining smooth transitions without sharp turns, suggesting steady-state laminar flow where the fluid moves smoothly like waves.
& \raisebox{-.9\height}{\includegraphics[width=0.12\textwidth]{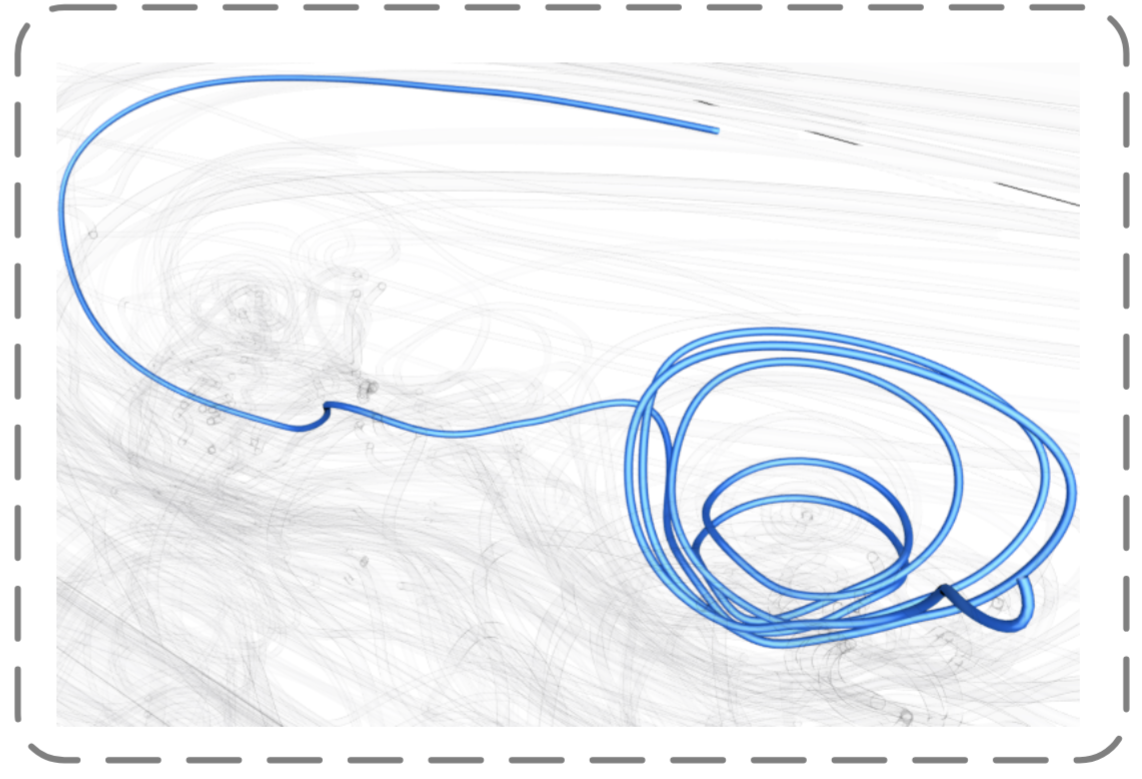}} \\

Illustrate the streamline.  
& The streamline pattern shows complex structures with \textbf{swirling, turbulent vortices} and regions of high shear stress caused by velocity gradients and fluid-obstacle interactions.  
& The flow pattern begins with smooth, curved lines. As it moves to the center, there is intricate turbulence creating complex but continuous patterns.
& \raisebox{-.9\height}{\includegraphics[width=0.12\textwidth]{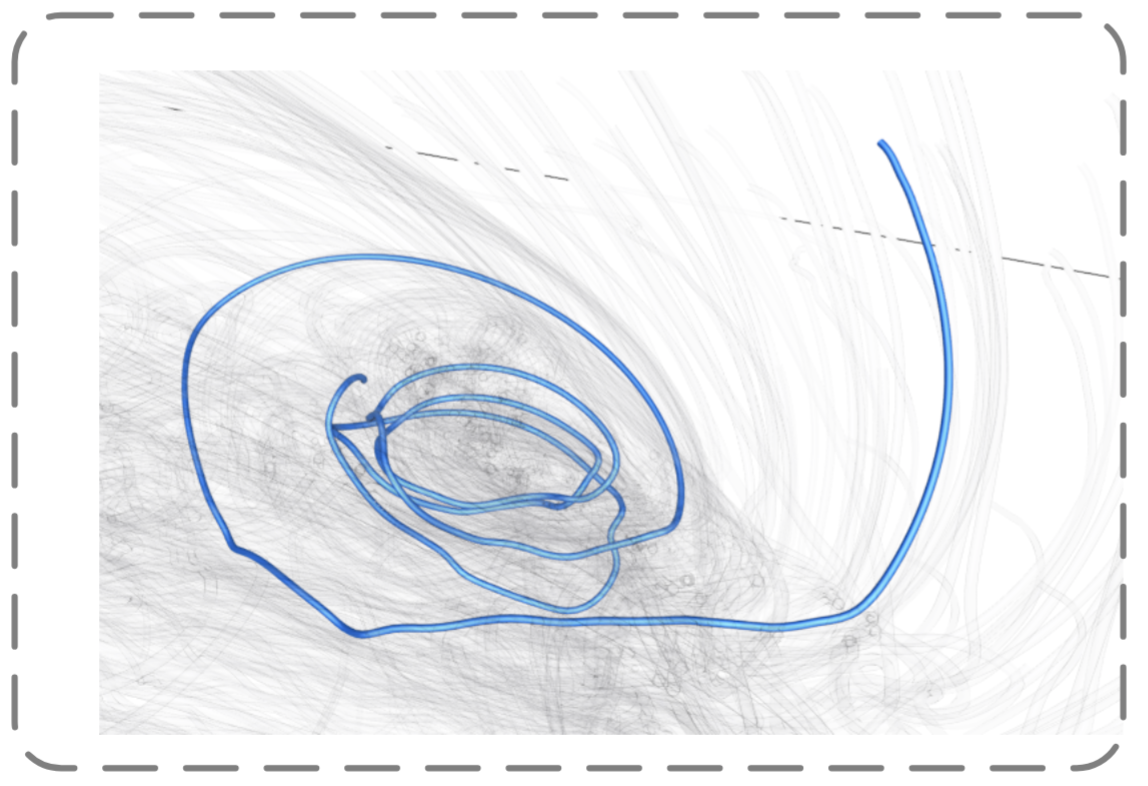}} \\

\bottomrule
\bottomrule
\end{tabular}
\end{table*}

\end{document}